\documentclass[fleqn,usenatbib,nofootinbib]{mnras}
\usepackage{graphicx}
\usepackage{amsmath,amssymb}
\usepackage{lastpage}
\usepackage[all]{hypcap}
\usepackage[T1]{fontenc}
\usepackage{float}
\usepackage{subfigure}
\usepackage{afterpage}
\usepackage[usenames]{color}
\usepackage{mathrsfs}
\usepackage{upgreek}
\usepackage{hyperref}
\usepackage{gensymb}
\usepackage{color}
\usepackage[dvipsnames]{xcolor}
\usepackage{longtable}
\usepackage{threeparttable}
\usepackage{multirow}
\usepackage{placeins}
\usepackage{orcidlink}
\usepackage{bm}
\usepackage[capitalise]{cleveref}
\usepackage{enumitem}

\title[Testing redshift-independent distance methods]{Consistencies and inconsistencies in redshift-independent distances}
\author[J. A. N\'ajera and H. Desmond]{
Jos\'e Antonio N\'ajera$^{\orcidlink{0000-0001-9738-7704}}$$^{1}$\thanks{E-mail: antonio.najera@port.ac.uk}
and Harry Desmond$^{\orcidlink{0000-0003-0685-9791}}$$^{1}$\thanks{E-mail: harry.desmond@port.ac.uk}
\\
$^{1}$Institute of Cosmology \& Gravitation, University of Portsmouth, Dennis Sciama Building, Portsmouth, PO1 3FX, UK\\\\
}

\pubyear{2025}

\begin{document}
\label{FirstPage}
\pagerange{\pageref{FirstPage}--\pageref{LastPage}}
\maketitle

\begin{abstract}
Redshift-independent distances underpin much of astrophysics, and there exists a plethora of methods to estimate them. However, the extent to which the distances they imply are consistent, while crucial for the integrity of the distance ladder, has been little explored. We construct a statistical framework to assess both internal (between measurements with the same method) and external (between-method) consistency by comparing differences between distances to their quoted
% $1\sigma$ C.L.
statistical uncertainties %\jan{Probably better to exclude the sentence that was after this since we don't use 10 of the indicators at all.} \hd{It seems alright to me as it is now} %We apply this framework to 76 different distance indicators to $>140,000$ galaxies \hd{do we actually use all of these in our analysis?} in the Nasa Extragalactic Distance Database (NED-D). We find that
% In this paper, we studied the Galaxy distance consistency in the Nasa Extragalactic Distance Database (NED-D) and SH0ES 2022 catalogues. NED-D gives redshift-independent measurement of distances for 76 different distance indicators. The consistency analyses involve taking all galaxies which have at least two measurements of a given indicator (internal) and two measurements of two different indicators (external) and taking all the possible pairs of distances. We computed the Mahalanobis distance for all pairs, build a distribution and compared it to the null hypothesis of a standard half-normal, which implies absence of systematics, outliers, and more problems. We studied the distance consistency with the Kullback-Leibler divergence (KLD). We simulated 10000 realizations and computed the $p$-value with the ratio of realizations with a higher KLD than the empirical case with the total realizations. We used a threshold of 0.05 to discard the null hypothesis. For NED-D,
in the NASA/IPAC Extragalactic Database of Distances (NED-D). 66 of the 76 indicators in NED-D are amenable to a consistency test by having at least two measurements to the same galaxy or at least one measurement to a galaxy also measured by another method.
% had at least one consistency analysis \hd{don't know what this means} and 7 showed perfect internal and external consistency.
We find that only 12 of these methods produce self-consistent distances across literature determinations, of which
7 are also consistent with distances to the same galaxies measured by all other methods.
% \jan{This sentence is important since it refers to the best set of indicators.}
The most consistent 6 methods (M-stars luminosity, Novae, Masers, Globular Cluster Fundamental Plane, O- and B-type Supergiants and BL Lac Luminosity) also give similar average distances to the mean of all indicators, while the 7th (Proper Motion) underestimates distances relative to the mean by 17.1\%. %\hd{quote the actual number}
% \hd{thought it was a factor $>2$?} \jan{it's a factor of 2 for the worse methods but not for this best set}.
We also investigate consistency of Cepheid distances in the SH0ES 2022 catalogue, finding no evidence for unaccounted-for systematics.
% showed consistency with the null hypothesis, having a $p$-value of $0.6613$, which is way above the threshold. This indicates that under this analysis, there is no evidence for unaccounted systematics in the catalogue. Thus, this implies no evidence of inconsistent means or uncertainties in the Cepheid-SNIa distance ladder as the cause of the Hubble-Lemaitre tension.
Our NED-D results imply that considerable work remains to obtain reliable distances by a multitude of methods, a crucial endeavour for constructing a multiply cross-checked  and fully robust distance ladder.
% , which implications for all inferences, such as measurement of the Hubble constant, that rely on this ladder.
We make our code publicly available.
\end{abstract}


\section{Introduction}
\label{sec:intro}

% Importance of redshift-independent distances
%
% Examples of that and what it leads to
%
% Importance of uncertainties in these and how that's determined kinda differently
%
% Relevant literature that we build on
%
% What we do and what it implies
%

% \hd{Make sure everything is in the present tense (the conclusion is the only sec where you can get away with past)} \jan{Partially done, I'm going over the manuscript to ensure this.}

% \hd{Write the priors for all MCMC analyses}
% \jan{Done}

% \hd{Might be nice to put the names of the methods in font like \texttt{this}. And only put them in bold if you want them to stand out because they're important methods in some way. Perhaps you should just put in bold the ones that are both internally consistent and in the maximal set of external consistency.} \jan{Done}

% \hd{Do you use Github? It would be good to put the code up there.} \jan{To be done}

% \hd{Some of the consistent methods have very few measurement pairs, which makes it easier for them to pass the consistency test (i.e. harder to acquire compelling evidence of inconsistency) and pretty useless in terms of building the distance ladder. We should make sure this is spelled out somewhere, and maybe highlight a few methods that are both fully consistent and might actually be useful.} \jan{This is expressed in paragraph 3 of 4.1}

Much of modern cosmology depends on redshift-independent distances to astrophysical objects. Historically the first application of such distances was to combine with redshifts to produce a ``Hubble diagram'', which provided the first evidence that the Universe is expanding~\citep{Hubble_1929} and that the expansion of the Universe is accelerating~\citep{Riess, Perlmutter}. Hubble diagrams remain a crucial tool in cosmology today, providing precise constraints on parameter values and distinguishing between different gravitational and cosmological models (e.g.~ \citealt{li2021hubble, pourojaghi2022can,  rezaei2022cosmographic, anton2024hubble}). This is particularly pertinent currently in the context of the ``Hubble tension'', a claimed $\sim$5$\sigma$ tension \citep{riess2022comprehensive} between the present-day expansion rate of the Universe, $H_0$, derived from the Hubble diagram versus inferred from the Cosmic Microwave Background (CMB) assuming the expansion history of the Universe implied by the standard model of cosmology, $\Lambda$CDM (for recent reviews see~\citealt{Freedman_Madore_Hubble, Riess_review}). Even more recently, the Hubble diagrams of baryon acoustic oscillations and Type Ia supernovae (SNIa) appear to favour dynamical dark energy models over a cosmological constant for explaining the Universe's accelerated expansion~\citep{DESI}.

Assuming a cosmological model (e.g. $\Lambda$CDM) with calibrated values for its free parameters, distances in the Universe may be estimated from objects' redshifts alone. However, within the local $\sim$100 Mpc redshifts provide inaccurate distance estimates because galaxies' velocities are not predominantly due to the expansion of the Universe. There is a significant contribution from ``peculiar velocities'', galaxies' individual motions superimposed on the Hubble flow generated by gravitational interaction with their cosmic environment (for reviews see~\citealt{Strauss_Willick, Turner_review}). 
% \purp{Another critical problem of using model-dependent distances is that it leads to circular reasoning by assuming the conclusion as a premise}.
If distances to such galaxies can be measured redshift-independently then peculiar and expansion velocities can be determined separately. The pattern of peculiar velocities provides evidence concerning the large-scale structure of the local Universe~\citep{Dupuy,Valade}, the growth rate of cosmic structure~\citep{Boruah, Said, Stiskalek}, extensions to General Relativity \citep{lyall2023testing, lyall2024constraining}, the large-scale isotropy of the Universe (e.g.~\citealt{Feindt,Watkins}) and the typicality or otherwise of the Milky Way's place in the Universe \citep{bovy2009galactic, mcmillan2016mass}.
It also allows the distance--redshift relation to test, rather than assume, cosmological models.

Given these many and varied opportunities, much effort has naturally been devoted to measuring redshift-independent (or direct) distances. This is done in a series of steps, likened to rungs of a ladder: direct geometric methods (mainly parallax) are used to establish distances to nearby objects (mainly stars within the Milky Way), which are in turn used to calibrate objects on the ``first rung'' of the distance ladder. The most common such objects are Cepheid stars---which obey a linear relation between their distance-independent pulsation period and distance-dependent luminosity (for reviews see~\citealt{Freedman_Madore,Anderson_Cepheids})---and the Tip-of-the-Red-Giant-Branch (TRGB) feature of the colour--magnitude diagram, which has a fixed luminosity arising from the
% and \purp{corresponds to the luminosity temperature location of the
Helium flash in an electron-degenerate stellar core of intermediate-to-low-mass stars. Hence, it acts as a standard candle (\citealt{trgb}; for a review see~\citealt{Li_Beaton}). Once calibrated, these methods can be used to establish direct distances to $\sim$50 Mpc. They in turn are used to calibrate the ladder's ``second rung''. A key method here is SNIa which are standardisable candles: after correction for the colour and width of their light-curves they have a fixed luminosity, so that if this is calibrated using e.g. Cepheids or TRGB the measured flux determines the distance. This allows the Hubble diagram to be extended out to $z \approx 1$ ($\sim$3 Gpc;~\citealt{Pantheon_plus,Union3,DES_Y5_SN}).

Parallax--Cepheid/TRGB--SNIa is the best-known route to the distance ladder, but there is in fact a plethora of methods for determining redshift-independent distances. Based on standard candle, standard ruler, standard siren and other techniques, these employ different objects, different physics, different measurement methods and different systematics, and operate at various locations of the ladder (i.e. requiring varying degrees of calibration). Given the crucial importance of the distance ladder for cosmology, and the tensions it has thrown up, it seems prudent to compare carefully as wide as possible a range of methods to assess possible systematic errors and with an eye to swapping out or supplementing some methods for others in constructing a maximally robust, precise and far-reaching ladder.

That is the aim of this paper. We take as our starting point the \textit{NASA/IPAC Extragalactic Database of Distances} (NED-D),\footnote{\url{https://ned.ipac.caltech.edu/Library/Distances/}} a compilation containing $>$270,000 extragalactic distances measured to $>$140,000 galaxies using 76 different distance indicators \citep{steer2016redshift}. Our particular focus here is on quantifying the statistical consistency of distances measured to the same galaxies, either of multiple estimates with the same method (self-consistency) or between methods (cross-consistency). This depends crucially on the quoted uncertainties, which are themselves model-dependent quantities with their own uncertainties.
% (at least when assumed to be Gaussian).
We apply the same methods to the Supernova $H_0$ for
the Equation of State (SH0ES) pipeline as a leading example of distance ladder-based $H_0$ inference framework \citep{riess2009redetermination, riess20113, riess2012cepheid, riess2022comprehensive}. Our approach will allow us to determine which methods appear most reliable, and whether systematics
% associated with misestimated uncertainties
may be biasing literature inferences employing the distance ladder. In case of inconsistency we can establish whether this is due to under- or over-estimated uncertainties and/or mean shifts in all distances inferred by some method. This provides a stepping stone towards construction of a more confidently reliable ladder employing multiple redshift-independent distance methods.

Our work builds on that of \citet{Singh}, who identified a systematic underestimation in uncertainties in NED-D that is most prominent for the most widespread methods and growing over time. This suggests that the precision of e.g. $H_0$ inferred from the distance ladder may be overstated, clearly of great importance to the aforementioned tension. Our improvements on this analysis are first to improve the statistical measure and test of consistency and second to perform an independent analysis of the SH0ES 2022 results \citep{riess2022comprehensive} as the first catalogue claiming a $5\sigma$ Hubble tension.
% between local measurements of $H_0$ using the distance ladder and Planck 2018 flat-$\Lambda$CDM constraints \citep{aghanim2020planck}.
This provides information on whether inconsistencies in distance measurements may be responsible for (some of) the tension.
% Thus, a statistical consistency analysis can shed some light on the possibility of the precision of $H_0$ being overstated and of the presence of unaccounted systematics. \\

The structure of the paper is as follows. In Sec.~\ref{sec:ObservationalData} we describe the NED-D and SH0ES data that we analyse. Sec.~\ref{sec:consistencyAnalysis} details our statistical methodology, while Sec.~\ref{sec:results} presents our findings. We discuss more general ramifications of the work as well as caveats in Sec.~\ref{sec:disc}, and conclude in Sec.~\ref{sec:conc}.

\section{Observational Data}
\label{sec:ObservationalData}

% We analysed the consistency of distance measurements in two catalogues: NED Redshift-Independent Distances (NED-D)
% \footnote{\url{https://ned.ipac.caltech.edu/Library/Distances/}}
% \citep{steer2016redshift} and one the latest releases from SH0ES\footnote{\url{https://github.com/PantheonPlusSH0ES/DataRelease}} \citep{riess2022comprehensive}.

\subsection{NED-D}

NED-D is an online compilation of redshift-independent extragalactic distances containing $>270,000$ distance measurements to $>140,000$ galaxies using 76 different methods \citep{steer2016redshift}. These methods, categorised as standard candles, standard rulers or ``secondary'' (defined as having a precision $\sim$20\% as opposed to the other ``primary'' methods with precision $\lesssim$10\%;~\citealt{steer2016redshift}), are as follows:
\begin{itemize}
    \item \textbf{Standard candles:} AGN time lag,  Asymptotic Giant Branch Stars (AGB), B-type Stars (B Stars), BL Lac Object Luminosity (BL Lac Luminosity), Black Hole, Blue Supergiant, Brightest Cluster Galaxy (BCG), Brightest Stars, Carbon Stars, Cepheids, Colour-Magnitude Diagrams (CMD), Delta Scuti, Flux-Weighted Gravity-Luminosity Relation (FGLR), Gamma-Ray Burst (GRB), Globular Cluster Luminosity Function (GCLF), Globular Cluster Surface Brightness Fluctuations (GC SBF), HII Luminosity Function (HII LF), Horizontal Branch, M Stars luminosity (M Stars), Miras, Novae, O- and B-type Supergiants (OB Stars), Planetary Nebula Luminosity Function (PNLF), Post-Asymptotic Giant Branch Stars (PAGB Stars), Quasar spectrum, RR Lyrae Stars, Red Clump, Red Supergiant Variables (RSV Stars), Red Variable Stars (RV Stars), S Doradus Stars, SNIa SDSS, SX Phoenicis Stars, Short Gamma-Ray Bursts (SGRB), Statistical, Subdwarf Fitting, Sunyaev-Zeldovich Effect (SZ effect), Surface Brightness Fluctuations (SBF), Tip of the Red Giant Branch (TRGB), Type II Cepheids, Type II Supernovae Radio (SNII radio), Type Ia Supernovae (SNIa), White Dwarfs, Wolf-Rayet, Gravitational Wave (Grav. Wave.)
    \item \textbf{Standard rulers:} CO ring diameter, Dwarf Galaxy Diameter, Eclipsing Binary, Globular Cluster Radii (GC radius), Grav. Stability Gas. Disk, Gravitational Lenses (G Lens), HII Region Diameters (HII), Jet Proper Motion, Masers, Orbital Mechanics (Orbital Mech.), Proper Motion, Ring Diameter, Type II Supernovae, Optical (SNII Optical).
    \item \textbf{Secondary methods:} D-Sigma, Diameter, Dwarf Ellipticals, Faber-Jackson, Fundamental Plane (FP), GC K vs. (J-K), GeV TeV ratio, Globular Cluster Fundamental Plane (GC FP), H I + optical distribution, Infra-Red Astronomical Satellite (IRAS), L(H{beta})-{sigma}, LSB galaxies, Magnitude, Mass Model, Radio Brightness, Sosies, Tertiary, Tully Estimate (Tully est), Tully-Fisher.
\end{itemize}

% The complete catalogue is available as a CSV file. The data is organized in several columns, for example "Galaxy ID" for the Host Galaxy, the distance modulus $m-M$, its error $\sigma_{m-M}$, and the distance indicator used. Furthermore,
More details on the physics of the methods can be found on the NED-D website.\footnote{\url{https://ned.ipac.caltech.edu/Library/Distances/distintro.html}} Some measurements are quoted for a specific value of $H_0$, in which case we rescale them to correspond to
% to compute the distance modulus. For these cases, we performed a re-scaling by changing this value to
$H_0 = 70 \text{km}/\text{s}/\text{Mpc}$. Others are quoted for a specific value of the distance modulus to the Large Magellanic Cloud (LMC), in which case we rescale them to $\mu_\text{LMC} = 18.50$.\footnote{\emph{Within} the set of methods calibrated against either $H_0$ or $\mu_\text{LMC}$ the choice of $70 \text{km}/\text{s}/\text{Mpc}$ and 18.5 is simply a convention and changing it will not affect differences between the extragalactic distance moduli. However this choice does affect relative distances \emph{between} methods calibrated with $H_0$ or $\mu_\text{LMC}$. Thus $H_0$ not being $70 \: \text{km}/\text{s}/\text{Mpc}$ or $\mu_\text{LMC}$ not being 18.5 may contribute to some of the inconsistencies we uncover.}
Other measurements come with neither an assumed $H_0$ nor an assumed $\mu_\text{LMC}$, in which case we cannot standardise them.
% \hd{Is there a potential miscalibration between distances tied to $H0$ vs $\mu_{LMC}$? I'm not sure why they would necessarily be put on the same footing as each other.} \jan{The  NASA HST Key Project selected $\mu_{LMC} = 18.50$ and thus, to place distances callibrated differently on equal footing, we need to standardize the calibration value to this selection.}
% For these cases, we also perform a re-scaling to treat all measurements on equal footing. We considered all the distance indicators and all the measurements, except the ones without a given error or with an error equal to zero.
We discard measurements without a quoted uncertainty, which cannot be statistically analysed. We consider all measurements to be independent. Various measures of average distance over multiple indicators in NED-D were defined and compared in~\citet{nedd_mean}. We denote distance indicators in NED-D with \texttt{this font}.

\subsection{SH0ES 2022}\label{sec:data_shoes}

We also consider the Cepheid catalogue used by the SH0ES collaboration in 2022. This is the latest full release of the SH0ES data. %\hd{Is there a reason we're using this older one? Is it the latest that's publicly available?}.
Furthermore, this is particularly pertinent as it has been used to argue for a $>5\sigma$ Hubble tension
% We considered this compilation to test the consistency of distances in a catalogue that was used to constrain the Hubble constant $H_0$. These results showed a $5\sigma$ confidence level (C.L.) tension
\citep{riess2022comprehensive}.
% with the flat-$\Lambda$CDM constraints from the Planck 2018 collaboration \citep{aghanim2020planck}. Thus, an analysis testing the consistency of different Cepheid distances to the same galaxy can shed light on the possibilities of unaccounted systematic errors. \\

The SH0ES 2022 catalogue includes 3130 Cepheid measurements across 40 galaxies. Two of them correspond to first-rung galaxies with direct geometric distances, one non-SNIa host galaxy, with the remainder in the second rung used to calibrate SNIa.
% and 38 are second rung galaxies with one non-SNIa host galaxy.
The third calibration galaxy, the Milky Way (MW), imposes an effective Gaussian prior on the Wesenheit absolute magnitude at a period of 10 days and solar metallicity $M^W_{H,1}$.
% \hd{What does this last phrase mean?} \jan{It means constraints on $M^W_{H,1}$ given by the MW Cepheids}.
The full catalogue is available in three FITS files. The data does not include the distance modulus $\mu$ to each Cepheid, but rather the period $P$, Wesenheit apparent magnitude $m^W_H$ in the F160W passband and excess metallicity $[O/H]$ along with their
% uncertainties and
covariance matrix. %\hd{of all variables or just metallicity?}
The absolute magnitude can be derived from the Period--Luminosity (PL) relation via \citep{riess2022comprehensive}
\begin{equation}
    \mu_0 = m^W_H - M^W_{H,1} - b_W (\log_{10} P - 1) - Z_W [O/H],
\end{equation}
where $M^W_{H,1}$ is the fiducial Wesenheit absolute magnitude at $P = 10$ days and solar metallicity, and $b_W$ and $Z_W$ are parameters defining the empirical relation between these variables.
% period, metallicity and absolute magnitude.

\citet{riess2022comprehensive} quote constraints for these parameters, which would enable us to compute the individual distance modulus for all Cepheids in the catalogue. However, it does not include the full covariance matrix necessary to derive their uncertainties. Thus, we ran a Monte Carlo Markov Chain (MCMC) analysis using the \texttt{emcee} sampler\footnote{\url{https://emcee.readthedocs.io/en/stable/}} \citep{foreman2013emcee} with 100 walkers %\hd{Explain why emcee and make clear which are the parameters being constrained}
and $>50$ times the autocorrelation length in samples to ensure convergence. The set of inferred parameters includes the distance modulus to 38 galaxies, the differences between the predicted and geometric distance moduli for the two first-rung galaxies (minimised to calibrate the PL relation), the PL relation parameters previously mentioned, the absolute magnitude of SNIa $M^0_B$, $\Delta zp$ (a parameter accounting for possible systematics associated with ground-based observations), and $5 \log H_0$.

Following SH0ES we use wide uniform priors in all parameters. The $\chi^2$ function is provided in Eq. 6 in \citep{riess2022comprehensive}. The contribution from the MW Cepheids is entered as two external constraints in $M^W_{H, 1}$ from the Hubble Space Telescope (HST) and Gaia EDR3. We do the inference by using the \texttt{emcee} code provided by \citet{riess2022comprehensive}. By doing this, we get their same constraints, including $H_0 = 73.04 \, \pm \, 1.01$ km/s/Mpc. This produces a fit to the distance modulus of each galaxy as an average over the Cepheids they contain. This is accurate as long as the individual Cepheids are statistically consistent, which we study both for the catalogue as a whole and each galaxy separately. %\hd{Did you check that you constraints on the parameters are consistent with what are quoted in the SH0ES paper?}
Finally, we get the covariance matrix for the parameters $\{M^W_{H,1}, b_W, Z_W \}$ over the chains, and used this result to derive the covariance matrix for the Cepheid distances using
\begin{equation}
\label{eqn:covariance}
    \text{Cov}(\mu_{0k}, \mu_{0l}) = \sum_{i=1}^M \sum_{j=1}^M \left(\dfrac{\partial \mu_{0k}}{\partial \theta_i}\right) \left(\dfrac{\partial \mu_{0l}}{\partial \theta_j}\right) \text{Cov}(\theta_i,\theta_j),
\end{equation}
where $\theta_{i,j}$ are the parameters $\{ m_{H, (k,l)}^W, M_{H,1}^W, b_W, Z_W \}$ ($M=4$). This ensures that the uncertainties are properly propagated into the distance moduli.
% Then, we are taking into account the covariance between two measurements of the Wesenheit relative magnitude and the covariance between the PL parameters.
With this result we can perform our consistency analysis on this catalogue. We carry this out on the whole catalogue and also on each galaxy separately to search for galaxy-specific systematics.

\section{Methodology}
\label{sec:consistencyAnalysis}

\subsection{Quantifying consistency}\label{sec:quan_cons}

We perform two kinds of consistency tests: internal (or self) and external (or cross). For the internal test we consider the measurements for a given method and a given galaxy with each other, while for the external test we compare for each galaxy the measurements from every pair of methods. This naturally restricts us to galaxies for which there are multiple measurements.

We compute the difference between each pair of measurements using
% in units of $\sigma$. For any pair of two measurements, we used
\begin{equation}
\label{eq:differenceSigma}
    \Delta_{ij} = \sqrt{(\mu_i-\mu_j)^T\:\Sigma_{ij}^{-1}\: (\mu_i-\mu_j)} \: ,
\end{equation}
%\hd{Changed notation to $\Delta$ because $\Delta \sigma$ leads to confusion with the $\sigma$ terms in $\Sigma$.}
where $\mu_{i,j}$ are the distance moduli, and $\Sigma$ is the covariance between the two measurements. In this case $\Sigma_{ij}$ is a number given by $\sigma_{\mu_i}^2+\sigma_{\mu_j}^2 - 2 \, \text{Cov}(\mu_i,\mu_j)$ where $\sigma_{\mu_{i,j}}$ are the diagonal uncertainty terms and $\text{Cov}(\mu_i,\mu_j)$ the cross-term between the two measurements. Note that $\text{Cov}(\mu_i,\mu_j)$ is assumed to be zero for all NED-D measurements, and hence only plays a role in the SH0ES analysis. Eq.~\ref{eq:differenceSigma} gives the Mahalanobis distance \citep{mai1936generalised} describing the difference in units of $\sigma$ between the measurements. It would be distributed as a standard half-normal if $\mu_i$ and $\mu_j$ were scattered from a common true value by their uncertainties (positive part of the standard normal $\mathcal{N}(\mu=0, \sigma^2 = 1)$ for $x \geq 0$).\footnote{Eq.~\ref{eq:differenceSigma} may be compared to eq. 1 of \citet{Singh},
% a different expression is used for the difference in units of sigma
$\Delta_{12} = |\mu_1-\mu_2|/\sigma_{\mu_1}$). This is not symmetric under $\mu_1 \leftrightarrow \mu_2$,
% and of comparing a random variable to a fixed number in its treatment of the uncertainties.
although the combination effectively used in that work, $(\Delta_{12} + \Delta_{21})/2$, is. Our formula has the justification of being derived from the convolution of two Gaussians assumed to be independent.
% \purp{Even so, the physically significant value by taking the mean of $\Delta_{12}$ over all pairs gives similar results with the multiple that the individual uncertainties should have to make the measurements consistent.}
}

We calculate the following quantities:
\begin{enumerate}
    \item For each method in the NED-D catalogue, we take all the galaxies having at least two measurements. We compute $\Delta_{ij}$ for all possible pairs across all galaxies in that method. This quantifies internal consistency.
    \item For all possible pairs of methods in the NED-D catalogue, we take all galaxies having at least one measurement by both methods. We compute $\Delta_{ij}$ for all possible pairs across all galaxies for this method pair. This quantifies external consistency.
    \item For the SH0ES 2022 catalogue, we compute $\Delta_{ij}$ for all possible measurement pairs across all galaxies. This is a global measure of consistency across the catalogue.
    \item For each galaxy individually in the SH0ES 2022 catalogue, we compute $\Delta_{ij}$ for all possible measurement pairs in this galaxy. This measures consistency for each galaxy separately.
\end{enumerate}
This produces in each case a $\Delta_{ij}$ distribution, which we then compare to the expected standard half-normal distribution. If the $\Delta_{ij}$ could not plausibly have been drawn from this distribution this implies an inconsistency, which could be due to unaccounted-for systematic errors, underestimated or overestimated uncertainties, outliers, inconsistent normalisations of distance and/or other problems. To assess this consistency we compare the $\Delta_{ij}$ and standard half-normal distributions using the
% \subsection{Kullback-Leibler (KL) divergence $p$-value}
% For a more generally effective method, we instead compare the real data to the standard half-normal using the
Kullback--Leibler divergence (KLD;~\citealt{kullback1951information, mackay2003information}).\footnote{One might think first to use the Kolmogorov--Smirnov (KS) test \citep{an1933sulla, smirnov1948table}. This is however unsuitable, as the $p$-value of a given test statistic depends on the size of the input array, here the number of measurement pairs. That this is not the number of independent measurements biases $p$ low, which can lead one to infer inconsistency even for mock data drawn from a standard half-normal when the sample size is large. This can be addressed in the case of self-consistency by computing the $p$-value using the number of independent measurements rather than the number of pairs, but it is not possible to assign unambiguously a number of effective samples in the cross-consistency test.} This is defined by
\begin{equation}
\label{eq:KLDivergence}
    D_{KL} (P\,| \,Q) = \int_{-\infty}^{\infty} dx \,  P(x) \log \left( \dfrac{P(x)}{Q(x)} \right),
\end{equation}
where $P$ is the empirical probability distribution function (pdf), $Q$ is the comparison pdf (standard half-normal). We validated our results by computing the KL divergence using the Simpson's rule and the entropy (given by $\sum_i P(x_i)  \log(P(x_i)/Q(x_i))$) and found that they give consistent comparable results with each other. %\hd{comparable to what? You haven't said how you do it fiducially yet. It's also not clear what using Simpson's rule and the entropy means in this context.}
Eq.~\ref{eq:KLDivergence} measures the similarity of the two pdfs. To convert the discrete measurements into a continuous KL pdf we apply Gaussian Kernel Density Estimation (KDE; \citealt{virtanen2020scipy}) with a bandwidth
% compute this probability, we need an estimation of the empirical pdf. For this purpose, we used the \texttt{scipy.stats.gaussian\_kde} Python function \citep{virtanen2020scipy}. This function approximates the pdf of a random variable in a non parametric way. The most important factor to take into account for this is the bandwidth. We considered one
of the form $h = C n^{-1/5}$ where $n$ is the number of measurement pairs. %\hd{You apply this to $P(x)$ before doing the integral don't you? Not to DKL.}
We then apply either Scott's \citep{scott2015multivariate} or Silverman's \citep{silverman2018density} method for setting the bandwidth, which we find to give near-identical results.
% factor for determining the bandwidth. We found that they gave very similar consistent results. \\
We also find similar results using a histogram (top-hat KDE) rather than Gaussian KDE.

This measures the ``distance'' between the measured distribution and standard half-normal, but does not by itself tell us how unexpected any such distance is. To determine this, we compute the KLD of 10,000 mock datasets generated assuming that the $\Delta_{ij}$ test statistic precisely follows a standard half-normal. This produces the KLD distribution corresponding to the null hypothesis that the measurements are perfectly consistent up to scatter described by their uncertainties.  This allows us to compute a $p$-value for rejection of the null hypothesis as the fraction of
% After computing the KL divergence, for each analysis, we took the number of measurements in each galaxy. Then, we generated this number of measurements by sampling from a Gaussian distribution. We computed the difference in units of sigma for this simulation and its KL divergence. This is called a realization. For each determination of the empirical KL divergence, we ran 10000 realizations. This gives us a distribution of the KL divergences that we expect to get from data coming from a theoretical standard half-normal distribution. We computed the $p$-value by taking the number of realizations
mock datasets having a KLD greater than the empirical one.
%This measures how extreme the data is compared to the expectation from the null hypothesis.
% us, the $p$-value provides a measurement of how extreme the empirical data compared to the simulated one under the null hypothesis that the empirical distribution comes from a standard half-normal distribution.
We use a threshold of 0.05 to reject the null hypothesis. If $p>0.05$ there is insufficient evidence to conclude that the measurements are not simply scattered by their uncertainty distributions; we term this ``consistency''. For the simulations to be on equal footing to the real data, we simulate the same number of measurement pairs as exists in NED-D. This ensures that the $p$-value is independent of the number of galaxies and measurements.
% discard the null hypothesis, i.e., there is not enough evidence to conclude that the empirical distribution does not come from a standard half-normal. \\

Our choice to define the null hypothesis as \emph{consistency} between the measurements means that significant evidence is required before concluding inconsistency. One could imagine instead defining the null hypothesis as \emph{inconsistency}, so that significant evidence would be required to conclude consistency. Our choice has two advantages. First, while ``consistency'' is a well-defined model (each galaxy has an underlying true distance, which produces the various measurements through scattering by the observational uncertainties), inconsistency is not: there is an infinity of ways in which two sets of measurements may be inconsistent. It is therefore difficult to see how a null hypothesis of inconsistency could be unambiguously defined. Second, our choice makes our results conservative in the sense that the inconsistencies that we identify are significant by construction. We caution however that while we apply the  term ``consistency'' to any comparison yielding $p>0.05$, we really mean only that the null hypothesis of consistency cannot be robustly ruled out. A $p$-value greater than 0.05 may still favour inconsistency. In some sense our results therefore provide a lower bound on the degree of inconsistency in the datasets we analyse.

% to mean of null hypothesis determines whether strong evidence The meaning of this term in this context is that if $p > 0.05$, there is not strong evidence to reject the null hypothesis, or no statistical significant tension with it. However, it is important to note that this is not the same as saying that if $p > 0.05$, then the distribution was sampled from a half-normal. Thus, whenever $p < 0.05$, we have strong evidence of problems like unaccounted-for systematics, outliers, non-Gaussian behaviour or others whereas $p > 0.05$ means not enough evidence for them}

Note that defining the $p$-value as the fraction \emph{greater} than the measured value corresponds to us doing a one-tailed hypothesis test, searching for inconsistency only in the direction that the real data is \emph{less} like a standard normal than the mocks. It is in fact possible for the data to be abnormal in the other direction, viz. \emph{more} like a standard normal than would be expected under the null hypothesis. This would be tested by a two-tailed test that calculates $p$ as the fraction of mock datasets \emph{more extreme} than the real data. We choose simply to flag such ``anomalously good'' cases, which correspond to the one-tailed $p$-value being $>0.975$.

\subsection{Extracting insight from inconsistency}
\label{sec:ins_inc}

The result of Sec.~\ref{sec:quan_cons} is a $p$-value for rejection of the null hypothesis of consistency between each method and any other (including itself) in the case of NED-D, and between all the Cepheid distances in the SH0ES catalogue either globally or galaxy-by-galaxy.

There are a number of insights we can draw from this. The first is to identify the ``best'' (most consistent) distance indicators in the NED-D catalogue. To achieve this, we order the methods in increasing external consistency fraction. For a given indicator, we consider all other indicators that share at least one galaxy. We define the external consistency fraction as the number of such pairs that are consistent divided by the total number. Then, we remove the method with the lowest percentage external consistency and again compute the external consistency percentage assuming that a discarded method is responsible for inconsistency in any pair containing it. We repeat this process until we remain with a set of methods that have complete external consistency with each other. This is the maximal set of externally consistent methods.

Next we can learn about the cause of the inconsistencies. To learn whether an inconsistency derives under- or overestimated uncertainties we perform an additional analysis by rescaling the distance uncertainties by the ratio of the uncertainty of each galaxy to the standard half-normal standard deviation ($\sigma'_i = (\Sigma/\sqrt{1-2/\pi}) \sigma_i $ where $\sigma'_i$ is the rescaled uncertainty of a given distance measurement, $\Sigma$ is the standard deviation of the empirical pdf  before the rescaling and $\sigma_i$ is the original uncertainty). If this %turns 
produces consistency, the original inconsistency must have been caused by the uncertainties. In that case, we then compare the variance of the $\Delta_{ij}$ distribution to that of the standard half-normal (1-2/$\pi$). If it is higher, then the uncertainties are underestimated, and vice versa.
% \hd{Even though shifted means isn't possible for self-consistency, surely self-inconsistency could also be caused by outliers or other problems? So the under vs overestimated uncertainties as calculated here might not actually be responsible for the inconsistency. We can test this by rescaling the uncertainties?} \jan{We'll do this soon.}

For the case of external consistency the situation is more complex as an inconsistency could derive from inconsistent means of the $\mu_i$ and $\mu_j$ distributions rather than anything to do with the uncertainties. To disentangle the possible causes, we repeat the consistency analysis with two modifications. First, for each galaxy in the method pair, we computed the mean distance modulus over the measurements of each method. We then shift the mean of one method onto that of the other and repeat the consistency test. If this transformation produces consistency, the original inconsistency must have been due to inconsistent means. The second modification is instead of shifting the means to rescale the uncertainties by the relative standard deviations of the $\mu$ distributions.
% Second, for each galaxy in the method pair, we computed the distance modulus overall standard deviations of both indicators. Then, we rescaled the uncertainties of one of the indicators by the ratio of these two overall standard deviations.
If this produces consistency, the original inconsistency must have come from the uncertainties. Note that having inconsistent means and uncertainties are not mutually exclusive. In some cases consistency may be achieved \emph{either} by shifting the means \emph{or} rescaling the uncertainties; in others doing both is required.
% It is possible for a given method pair to have both, in which case neither of the above modifications will produce consistency by itself, but combining both of them will yield an acceptable $p$-value.
If the test returns inconsistency even after both of these are done, 
% even in this case the test returns inconsistency, this
it must be due to other effects such as outliers, non-Gaussian uncertainties or systematic errors. We call this an ``essential inconsistency,'' indicating that it cannot (easily) be removed.

\begin{table*}
    \centering
        \caption{Results for the standard candle distance indicators. We present the number of galaxies having at least two measurements by a given method, the number of measurement pairs in these galaxies, the total number of measurements in NED-D along with the number of papers from which these measurements derive, the KLD $p$-value and external consistency statistics. The $p$-value is shown in green if it is greater than 0.05 (indicating consistency, i.e. not enough evidence to reject the null hypothesis), orange if it is between 0 and 0.05 (indicating inconsistency), and red if it is 0 (indicating severe inconsistency). In cases of inconsistency, we indicate the $p$-value after rescaling the uncertainties to a standard half-normal in parenthesis. If this produces consistency, we indicate whether the uncertainties were originally underestimated (downward-pointing arrow) or overestimated (upward-pointing arrow). The final column shows the fraction of other methods with which the method in question is externally consistent: the first number gives the number of cases that are consistent out-of-the-box, the second the number that are made consistent by shifting the means but not by rescaling the uncertainties, the third the number that are made consistent by rescaling the uncertainties but not by shifting the means, the fourth the number that are made consistent by \emph{either} shifting the means \emph{or} rescaling the uncertainties, and the fifth the number that are made consistent only by \emph{both} shifting the means \emph{and} rescaling the uncertainties. The final number shows the total number of other methods with which a given method is compared, such that this number minus the sum of the others is the number of essential inconsistencies.} Methods that are internally consistent and members of the maximal externally consistent set are shown in bold. (See text for full details.) %\hd{Can you put spaces around the slashes?}
        
    \label{tab:results-standard-candles}
    \begin{tabular}{cccccc}
    \hline
    Distance indicator & \# of galaxies & \# pairs & \# total measurements/references & KL $p$-value & External consistency \\
    \hline
\textbf{BL Lac Luminosity} & 2 & 4 & 20 / 11 & \textcolor{OliveGreen}{0.9705} & 1 - 0 - 0 - 0 - 0 / 1 \\ 
RSV Stars & 2 & 4 & 9 / 6 & \textcolor{OliveGreen}{0.8298} & 9 - 0 - 3 - 4 - 0 / 28 \\ 
\textbf{OB Stars} & 2 & 2 & 4 / 3 & \textcolor{OliveGreen}{0.7383} & 27 - 0 - 1 - 0 - 0 / 31 \\ 
\textbf{Novae} & 6 & 14 & 18 / 11 & \textcolor{OliveGreen}{0.4382} & 35 - 0 - 0 - 3 - 1 / 46 \\ 
Miras & 6 & 119 & 36 / 27 & \textcolor{OliveGreen}{0.1121} & 20 - 0 - 4 - 2 - 0 / 39 \\ 
RV Stars & 1 & 10 & 5 / 1 & \textcolor{OliveGreen}{0.0845} & 6 - 0 - 1 - 0 - 0 / 18 \\ 
Statistical & 14 & 168 & 292 / 27 & \textcolor{OliveGreen}{0.0795} & 15 - 0 - 10 - 3 - 0 / 54 \\ 
Brightest Stars & 21 & 127 & 129 / 41 & \textcolor{OliveGreen}{0.074} & 21 - 0 - 6 - 3 - 2 / 44 \\ 
\textbf{M Stars} & 1 & 3 & 6 / 5 & \textcolor{OliveGreen}{0.0686} & 18 - 0 - 1 - 1 - 0 / 33 \\ 
Delta Scuti & 2 & 29 & 12 / 2 & \textcolor{RedOrange}{0.0395} (\textcolor{OliveGreen}{0.8241}) $\uparrow$ & 14 - 0 - 4 - 1 - 0 / 31 \\ 
Horizontal Branch & 22 & 139 & 110 / 53 & \textcolor{RedOrange}{0.0037} (\textcolor{OliveGreen}{0.1495}) $\downarrow$ & 19 - 0 - 4 - 4 - 0 / 40 \\ 
PNLF & 45 & 285 & 183 / 57 & \textcolor{RedOrange}{0.0016} (\textcolor{RedOrange}{0.0019}) & 23 - 0 - 12 - 3 - 1 / 52 \\ 
AGN time lag & 11 & 13 & 32 / 3 & \textcolor{BrickRed}{0.0} (\textcolor{OliveGreen}{0.7099}) $\downarrow$ & 2 - 0 - 3 - 1 - 0 / 6 \\ 
Carbon Stars & 7 & 182 & 82 / 13 & \textcolor{BrickRed}{0.0} (\textcolor{OliveGreen}{0.4555}) $\uparrow$ & 25 - 0 - 1 - 1 - 0 / 38 \\ 
Cepheids & 115 & 69496 & 10956 / 325 & \textcolor{BrickRed}{0.0} (\textcolor{BrickRed}{0.0}) & 8 - 0 - 6 - 2 - 1 / 47 \\ 
CMD & 123 & 41733 & 1265 / 170 & \textcolor{BrickRed}{0.0} (\textcolor{BrickRed}{0.0}) & 15 - 0 - 6 - 5 - 1 / 51 \\ 
GRB & 139 & 1661 & 729 / 17 & \textcolor{BrickRed}{0.0} (\textcolor{BrickRed}{0.0}) & - \\ 
GCLF & 176 & 2288 & 781 / 96 & \textcolor{BrickRed}{0.0} (\textcolor{BrickRed}{0.0}) & 15 - 0 - 4 - 1 - 0 / 40 \\ 
RR Lyrae & 2443 & 16262 & 26409 / 216 & \textcolor{BrickRed}{0.0} (\textcolor{BrickRed}{0.0}) & 16 - 1 - 4 - 2 - 2 / 43 \\ 
Red Clump & 24 & 3420 & 236 / 63 & \textcolor{BrickRed}{0.0} (\textcolor{BrickRed}{0.0}) & 16 - 0 - 4 - 1 - 1 / 40 \\ 
SNIa SDSS & 1849 & 7687 & 6054 / 1 & \textcolor{BrickRed}{0.0} (\textcolor{BrickRed}{0.0}) & 0 - 0 - 2 - 0 - 0 / 4 \\ 
SZ effect & 78 & 426 & 283 / 19 & \textcolor{BrickRed}{0.0} (\textcolor{BrickRed}{0.0}) & 6 - 0 - 0 - 1 - 0 / 16 \\ 
SBF & 406 & 4403 & 1781 / 67 & \textcolor{BrickRed}{0.0} (\textcolor{BrickRed}{0.0}) & 19 - 0 - 8 - 2 - 1 / 42 \\ 
TRGB & 307 & 8537 & 1736 / 336 & \textcolor{BrickRed}{0.0} (\textcolor{BrickRed}{0.0}) & 15 - 1 - 3 - 9 - 0 / 54 \\ 
Type II Cepheids & 7 & 435 & 62 / 22 & \textcolor{BrickRed}{0.0} (\textcolor{OliveGreen}{0.3613}) $\downarrow$ & 18 - 1 - 6 - 1 - 0 / 44 \\ 
SNIa & 4978 & 129581 & 30843 / 140 & \textcolor{BrickRed}{0.0} (\textcolor{BrickRed}{0.0}) & 10 - 0 - 5 - 2 - 0 / 38 \\ 
AGB & 0 & 0 & 2 / 2 & - & 22 - 0 - 0 - 0 - 0 / 23 \\ 
B Stars & 0 & 0 & 2 / 2 & - & 28 - 0 - 0 - 0 - 0 / 30 \\ 
Black Hole & 0 & 0 & 17 / 2 & - & 4 - 1 - 2 - 1 - 0 / 11 \\ 
Blue Supergiant & 0 & 0 & 2 / 1 & - & 11 - 0 - 0 - 0 - 0 / 25 \\ 
BCG & 0 & 0 & 239 / 1 & - & 9 - 0 - 2 - 0 - 0 / 11 \\ 
FGLR & 0 & 0 & 7 / 7 & - & 9 - 0 - 3 - 4 - 0 / 21 \\ 
GC SBF & 0 & 0 & 3 / 1 & - & 13 - 0 - 2 - 0 - 0 / 19 \\ 
HII LF & 0 & 0 & 17 / 2 & - & 6 - 0 - 0 - 1 - 0 / 15 \\ 
SX Phe Stars & 0 & 0 & 3 / 2 & - & 10 - 0 - 0 - 0 - 0 / 13 \\ 
Subdwarf fitting & 0 & 0 & 1 / 1 & - & 19 - 0 - 0 - 0 - 0 / 22 \\ 
SNII radio & 0 & 0 & 24 / 2 & - & 3 - 1 - 6 - 4 - 0 / 17 \\ 
White Dwarfs & 0 & 0 & 1 / 1 & - & 21 - 0 - 0 - 0 - 0 / 22 \\ 
Wolf-Rayet & 0 & 0 & 1 / 1 & - & 0 - 0 - 1 - 3 - 0 / 5 \\ 
Grav. Wave & 0 & 0 & 8 / 8 & - & 4 - 0 - 0 - 0 - 0 / 4 \\ 
\hline

    \end{tabular}

\end{table*}

\begin{table*}
    \centering
    \caption{As Table~\ref{tab:results-standard-candles}, but for standard ruler (above the horizontal line) and secondary (below the line) distance methods.}
    \label{tab:results-standard-rulers}
    \begin{tabular}{ccccccc}
    \hline
    Distance indicator & \# of galaxies & \# pairs & \# total measurements/references & KL $p$-value & External consistency \\ \hline
\textbf{Proper Motion} & 1 & 6 & 5 / 3 & \textcolor{OliveGreen}{0.8918} & 24 - 2 - 3 - 0 - 0 / 32 \\ 
\textbf{Maser} & 6 & 29 & 24 / 20 & \textcolor{OliveGreen}{0.6253} & 28 - 0 - 4 - 2 - 0 / 36 \\ 
Dwarf Galaxy Diameter & 8 & 8 & 16 / 1 & \textcolor{RedOrange}{0.0004} (\textcolor{OliveGreen}{0.582}) $\uparrow$ & 17 - 0 - 3 - 0 - 0 / 21 \\ 
G Lens & 15 & 70 & 86 / 14 & \textcolor{RedOrange}{0.0002} (\textcolor{OliveGreen}{0.6975}) $\downarrow$ & 1 - 0 - 2 - 0 - 0 / 3 \\ 
Eclipsing Binary & 49 & 3397 & 273 / 48 & \textcolor{BrickRed}{0.0} (\textcolor{RedOrange}{0.0007}) & 22 - 0 - 2 - 1 - 0 / 42 \\ 
HII region diameter & 19 & 25 & 53 / 7 & \textcolor{BrickRed}{0.0} (\textcolor{RedOrange}{0.0261}) & 7 - 0 - 15 - 7 - 0 / 42 \\ 
Ring Diameter & 47 & 47 & 212 / 1 & \textcolor{BrickRed}{0.0} (\textcolor{OliveGreen}{0.2025}) $\uparrow$ & 10 - 0 - 6 - 1 - 0 / 18 \\ 
SNII optical & 309 & 6090 & 1622 / 89 & \textcolor{BrickRed}{0.0} (\textcolor{BrickRed}{0.0}) & 17 - 0 - 4 - 3 - 0 / 42 \\ 
GC radius & 0 & 0 & 101 / 8 & - & 22 - 0 - 9 - 2 - 0 / 37 \\ 
Grav. Stability Gas. Disk & 0 & 0 & 2 / 1 & - & 1 - 0 - 0 - 0 - 0 / 4 \\ 
Orbital Mech. & 0 & 0 & 3 / 3 & - & 20 - 0 - 0 - 2 - 0 / 22 \\       \hline
\textbf{GC FP} & 1 & 3 & 3 / 3 & \textcolor{OliveGreen}{0.2469} & 16 - 1 - 0 - 0 - 0 / 19 \\ 
GeV TeV ratio & 4 & 8 & 30 / 11 & \textcolor{RedOrange}{0.0385} (\textcolor{OliveGreen}{0.7002}) $\downarrow$ & 1 - 0 - 0 - 0 - 0 / 1 \\ 
Faber-Jackson & 427 & 1947 & 1424 / 6 & \textcolor{RedOrange}{0.0101} (\textcolor{OliveGreen}{0.6445}) $\uparrow$ & 5 - 0 - 11 - 5 - 0 / 29 \\ 
Diameter & 3 & 3 & 6 / 2 & \textcolor{RedOrange}{0.0089} (\textcolor{OliveGreen}{0.0504}) $\uparrow$ & 18 - 0 - 2 - 0 - 0 / 31 \\ 
D-Sigma & 551 & 2901 & 1995 / 9 & \textcolor{BrickRed}{0.0} (\textcolor{BrickRed}{0.0}) & 9 - 0 - 4 - 1 - 0 / 25 \\ 
FP & 967 & 1675 & 130214 / 22 & \textcolor{BrickRed}{0.0} (\textcolor{BrickRed}{0.0}) & 10 - 0 - 5 - 2 - 0 / 28 \\ 
IRAS & 405 & 640 & 2947 / 2 & \textcolor{BrickRed}{0.0} (\textcolor{BrickRed}{0.0}) & 4 - 1 - 16 - 0 - 0 / 35 \\ 
Mass Model & 2 & 4 & 7 / 7 & \textcolor{BrickRed}{0.0} (\textcolor{OliveGreen}{0.8942}) $\downarrow$ & 0 - 0 - 0 - 0 - 0 / 10 \\ 
Tully-Fisher & 9107 & 231663 & 56436 / 78 & \textcolor{BrickRed}{0.0} (\textcolor{BrickRed}{0.0}) & 13 - 1 - 5 - 3 - 0 / 48 \\ 
H I + optical distribution & 0 & 0 & 1 / 1 & - & 0 - 0 - 0 - 0 - 0 / 1 \\ 
LSB galaxies & 0 & 0 & 2 / 1 & - & 7 - 0 - 0 - 1 - 0 / 10 \\ 
Magnitude & 0 & 0 & 111 / 3 & - & 13 - 0 - 5 - 12 - 0 / 40 \\ 
Sosies & 0 & 0 & 287 / 3 & - & 10 - 0 - 7 - 1 - 0 / 21 \\ 
Tertiary & 0 & 0 & 284 / 4 & - & 6 - 0 - 3 - 2 - 0 / 22 \\ 
Tully est & 0 & 0 & 1434 / 1 & - & 17 - 0 - 8 - 0 - 0 / 41 \\     \hline
    \end{tabular}

\end{table*}

We are also interested in studying the tendency of each method to produce smaller or larger distances relative to the others. This will help to determine the methods that under- or overestimate distances with respect to other methods. For any given method pair and galaxy, we calculate the uncertainty-weighted mean distance modulus as
\begin{equation}
\bar{\mu}^i_g = \frac{\sum_k \mu^i_{g,k}/(\sigma^i_{g,k})^2}{\sum_k 1/(\sigma^i_{g,k})^2}
\end{equation}
and its uncertainty $\sigma^i_g = 1/\sqrt{\sum_k {(\sigma^i_{g,k}})^{-2}}$ where $i$ represent the indicator, $g$ the galaxy and $k$ goes over all the measurements for this indicator and galaxy.  %\hd{is this the standard error on the mean, or what?}
We get these values for both methods. Then, we compute the uncertainty-weighted mean across all galaxies
\begin{equation}
\bar{\mu}_i = \frac{\sum_g \bar{\mu}^i_g/(\sigma^i_g)^2}{\sum_g 1/(\sigma^i_g)^{2}}
\end{equation}
and its uncertainty $\sigma_{\bar{\mu}_i} = 1/\sqrt{\sum_g (\sigma^i_g)^{-2}}$. %\hd{don't understand this sentence}.
We then construct a difference $X_{ij} \equiv \bar{\mu}_i - \bar{\mu}_j$, %\hd{isn't it $mu$?}
% where $\bar{\mu}_i$ is the average of weighted means for the $i$-th distance indicator across all galaxies. %\hd{this is all for one galaxy, right?}.
with uncertainty $\sigma_{X_{ij}} = \sqrt{\sigma_{\bar{\mu}_i}^2 + \sigma^2_{\bar{\mu}_j}}$.
% with $\sigma_{\bar{D}_{i,j}}$ the uncertainties of the average of weighted means.
% We build a set of equations
% \begin{equation}
%     \bar{\mu}_i - \bar{\mu}_j = X_{ij},
% \end{equation}
%\hd{We should call this $\bar{\mu}$ or something to indicate that it's some kind of average rather than being a specific measurement through method $i$.}
This defines $X_{ij}$ for all methods with at least one galaxy in common, which we
% for all the distance indicator pairs having at least one measurement pair in at least one common galaxy. %\hd{For a given method pair, there is one equation like this for each galaxy they have in common?}.
then use as constraints to infer the $\bar{\mu}_i$.
% This set of equations, in general, does not need to follow the consistency relations given by expressions of $X_{ij}$. Thus, the system will not have an exact solution. However, we can get the set of $\mu_i$ that minimizes the residual $\sum_{ij}(X_{ij \, \text{pred}} - X_{ij})^2/\sigma_{X_{ij}}^2$ with $X_{ij \, \text{pred}} = \bar{\mu}_{i \, \text{pred}} - \bar{\mu}_{j \, \text{pred}}$.
Since not all method pairs have an $X_{ij}$ some methods will have more constraints than others, leading to better-determined $\bar{\mu}_i$.
% Furthermore, since not all the method pairs will have an equation, some methods will have more constraints than others.
There is even the possibility of having methods with zero constraints or only one constraint. The former would imply that we cannot determine $\bar{\mu}_i$ for that method while the latter can cause degeneracies in the $\bar{\mu}_i$ of different indicators. It is important to realise that $\bar{\mu}_i$ represents an average relative distance inferred by method $i$, which must therefore be defined relative to an arbitrary zero-point. %\hd{even then they're not actual distances. It's best to say that these are relative distances, which we define wrt the mean over all methods}.
We take this to be the mean of all the $\bar{\mu}$.
% Thus, the constraints on $\bar{\mu}_i$ gives us information on the overall distances of each indicator relative to the mean of all. The choice of the zero point is arbitrary and we selected this for an easier interpretation of the data.
We constrain the $\bar{\mu}_i$ using the NUTS method of Hamiltonian Monte Carlo as implemented in the NumPyro sampler \citep{bingham2019pyro, phan2019composable}, with wide uniform priors on the $\bar{\mu}_i$, ensuring convergence through the Gelman--Rubin statistic. %\hd{describe numpyro like you did emcee}

\section{Results}
\label{sec:results}

\subsection{NED-D}\label{sec:results_ned}

From the 76 indicators in NED-D, only 66 have at least one galaxy containing either two measurements with that indicator or one with that indicator and one with another. The others, which we cannot analyse for consistency and hence neglect for the remainder of the paper, are \texttt{Quasar spectrum, SGRB, PAGB Stars, S Doradus Stars, GC K vs. (J-K), CO ring diameter, Radio Brightness, L(H\{beta\})-\{sigma\}, Dwarf Ellipticals and Jet Proper Motion}.

\begin{figure*}
    \centering
    \includegraphics[width=0.9\linewidth]{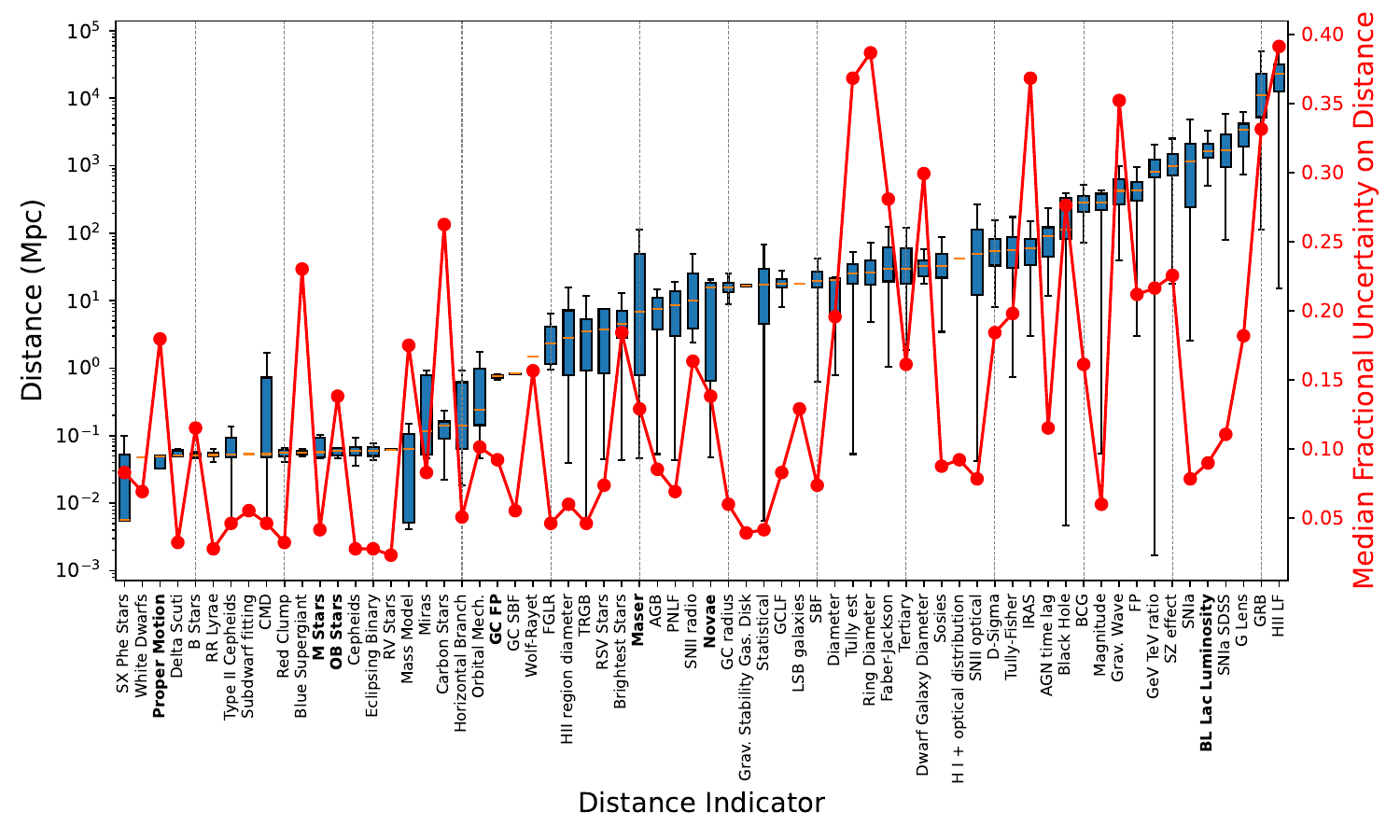}
    \caption{Whisker plot showing the full range (whiskers) and interquartile range (boxes) of the distances probed by each indicator in NED-D. The median distances are shown by orange bars.
    % We exclude the outliers using the Interquartile method (we exclude the distances having a gap higher than $1.5 \, (Q_3-Q_1)$ to the $Q_1$ first and $Q_3$ third quartiles). 
    We also show the median fractional distance uncertainty of each method in red, with scale on the right-hand y-axis.
    }
    \label{fig:whiskerPlotDistances}
\end{figure*}

To set the stage we show in Fig.~\ref{fig:whiskerPlotDistances} the distributions of distances probed by each method we consider. For each method the box shows the median distance (orange bar) and interquartile range, while the whisker covers the full range. The methods are ordered in median distance.
We can see that 17 indicators (from \texttt{SX Phe Stars} to \texttt{Mass Model}) measure median distances lower than $0.1$ Mpc. Then, we have seven indicators (from \texttt{Miras} to \texttt{GC SBF}) having distances from 0.1 Mpc to 1 Mpc and nine from 1 Mpc to 10 Mpc (\texttt{Wolf-Rayet} to \texttt{PNLF}). These three ranges represent the first rung in the distance ladder. The second rung goes from a few Mpc to around several tens of Mpc (\texttt{Wolf-Rayet} to \texttt{Black Hole}). The third rung covers $\sim$100 Mpc to $\sim$400 Mpc. The SNIa method extends to thousands of Mpc, while \texttt{GRB} and \texttt{HII LF} reach even beyond 10 Gpc albeit with large relative uncertainties ($\gtrsim30\%$).
% are interested in visualizing the distance indicators having the highest and lowest measures of extragalactic distances. For this, we can build a whisker plot showing the median, interquartile ranges and full ranges of distances in megaparsecs (Mpc). We can exclude the distance outliers using the interquartile method. Given $Q_1$ the first and $Q_3$ the third quartile, the interquartile range is given by $IQR = Q_3-Q_1$. Then, the interquartile method for exclusion of outliers considers elements between $Q_1 - 1.5 \, IQR$ and $Q_3 + 1.5 \, IQR$. The second quartile corresponds to the median $Q_2$. We are also interested in which indicators have higher fractional uncertainties on the distance.
We also show in red the median fractional uncertainty of all measurements using a given indicator in NED-D, with scale on the right-hand y-axis.
The methods highlighted in bold are those that we will find to be most consistent. Most of these have median relative errors lower than 20\%, and two even below 10\% (\textbf{M Stars} and \textbf{BL Lac Luminosity}). \textbf{BL Lac Luminosity} also probes to several Gpc.
To be of maximum usefulness in the distance ladder, an indicator would ideally be self-consistent and consistent with other (good) methods, and probe large distances with small uncertainty.
% \purp{Finally, Fig, \ref{fig:whiskerPlotDistances} presents a whisker plot showing the full range as the whiskers and the interquartile range as boxes. We exclude the outliers with the interquartile method previously described. We show in red the median of the fractional uncertainty on the distance to see what are the indicators having higher relative uncertainties.

Our consistency results are summarised in Tables \ref{tab:results-standard-candles} for the standard candle distance indicators and \ref{tab:results-standard-rulers} for the standard ruler and secondary indicators.  We report, for each method, the number of galaxies that have at least two distance measurements using the given indicator, the number of measurement pairs for these galaxies, the total number of measurements in the NED-D catalogue from this indicator and the number of references reporting these measurements, the KLD $p$-value (with a traffic-light-style colouring indicating the severity of the inconsistency) and external consistency statistics. For self-inconsistent cases we show the $p$-value after rescaling the uncertainties to give $\Delta_{ij}$ the variance of a standard half-normal distribution in parentheses. The final column contains the number of other distance indicators with which the method in question is consistent ($p>0.05$), in five different hyphen-separated cases. The first number describes the fiducial case with no shifting of means or rescaling of uncertainties. The second number gives the number of pairs that are made consistent when the means of the $\mu$ distributions are made to coincide, thus eliminating discrepant central values as a cause of inconsistency. The third number is the same but for rescaling the uncertainties so that the $\Delta_{ij}$ distribution has the width of a standard half-normal by construction, eliminating discrepant uncertainties as a cause of uncertainty. The fourth number corresponds to the case in which either shifting the means or rescaled the uncertainties produces consistency, and the fifth to the case where doing both is required to produce consistency.
% , eliminating both of these as possible causes. %\hd{Shouldn't there be another one for both shifted mean and rescaled uncertainty?}
The final number is the total number of other methods with which that method has overlap, i.e. the largest number of pairs that could possibly be consistent.

Tables \ref{tab:results-standard-candles} and \ref{tab:results-standard-rulers} show that most of the NED-D distance indicators are internally inconsistent, largely for an unidentified reason. This indicates that the catalogued distances are subject to problems like systematics and statistical outliers and are likely unsuitable for precision analyses such as constraining the Hubble constant. Indeed, the methods normally used for inferring $H_0$---Cepheids, TRGB and SNIa---are among the methods with the highest number of measurement pairs and all have a consistency $p$-value of 0. We also show the results for only the measurements that can be standardised, i.e. the ones with a quoted value for either $H_0$ or $\mu_{LMC}$. These results are shown in Appendix \ref{sec:app} (Figs.~\ref{tab:results-standard-candles-calibrated} and \ref{tab:results-standard-rulers-calibrated}), where we see for example that the \texttt{CMD} method becomes much better behaved.

\begin{figure*}
    \centering
    \includegraphics[width=1.05\linewidth]{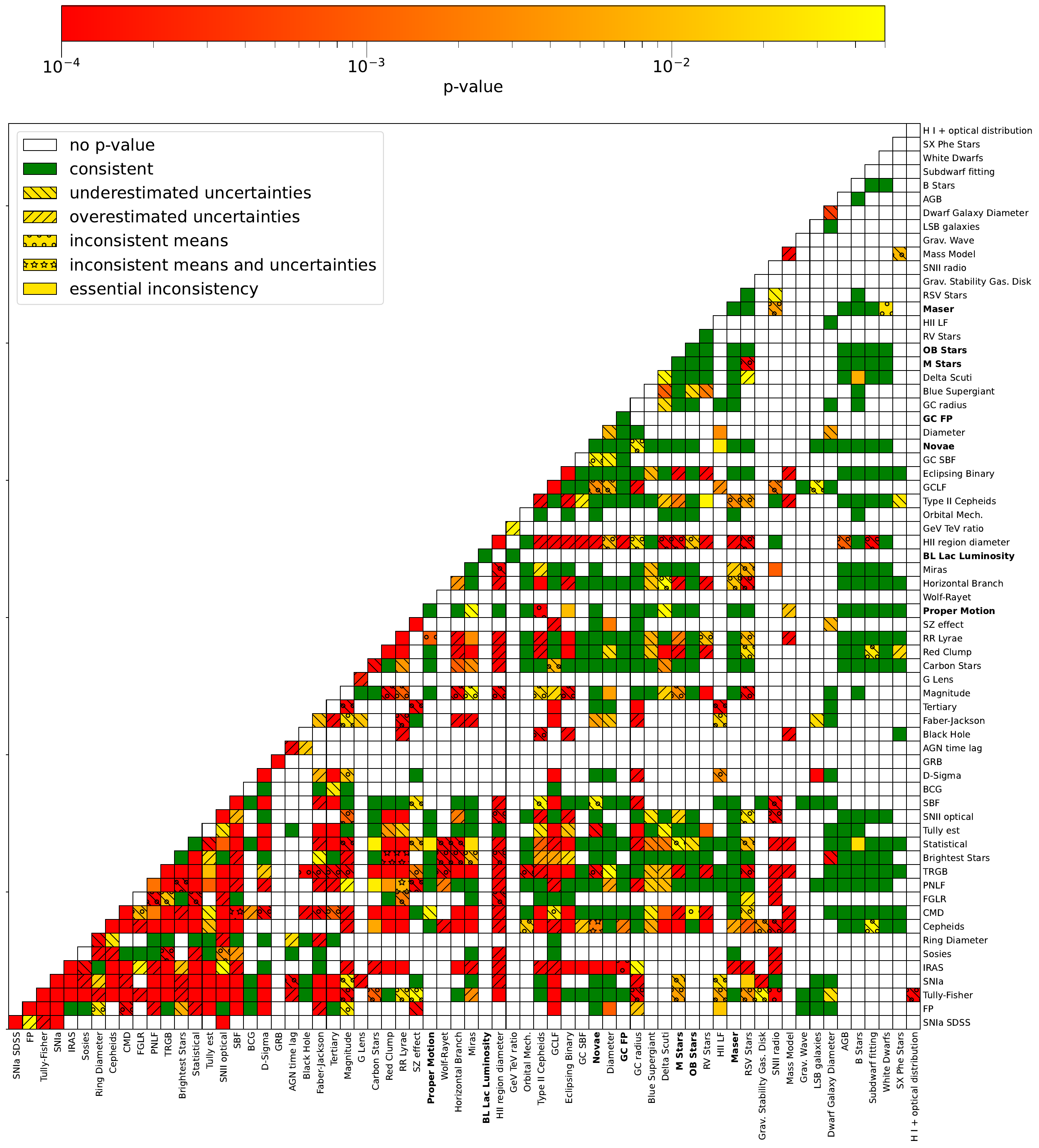}
    \caption{Consistency $p$-value results for the NED-D distance indicators. The diagonal squares describe internal consistency whereas the non-diagonal squares describe external consistency with other methods. We show the cases with $p$-value higher than 0.05 (i.e. not rejecting the null hypothesis) in green, the cases with no reported $p$-value in white and the cases with a $p$-value lower than 0.05 in a logarithmic colour scale. We show cases with underestimated uncertainties with a backslash hatching, overestimated uncertainties with forward-slash hatching, and inconsistent means with circle hatching. If either shifting the means or rescaling the uncertainties produces consistency we show both types of hatching, while if both are required we show star hatching. If the cause of the inconsistency is neither one of these nor their combination (i.e. it remains inconsistent after shifting the means and rescaling the uncertainties) we show no hatching. We call this an essential inconsistency. %\hd{Could you align the right-hand edge with the final box? There seems to be part of another box visible.}
    }
    \label{fig:$p$-values}
\end{figure*}

\begin{figure}
    \centering
    \includegraphics[width=1.0\linewidth]{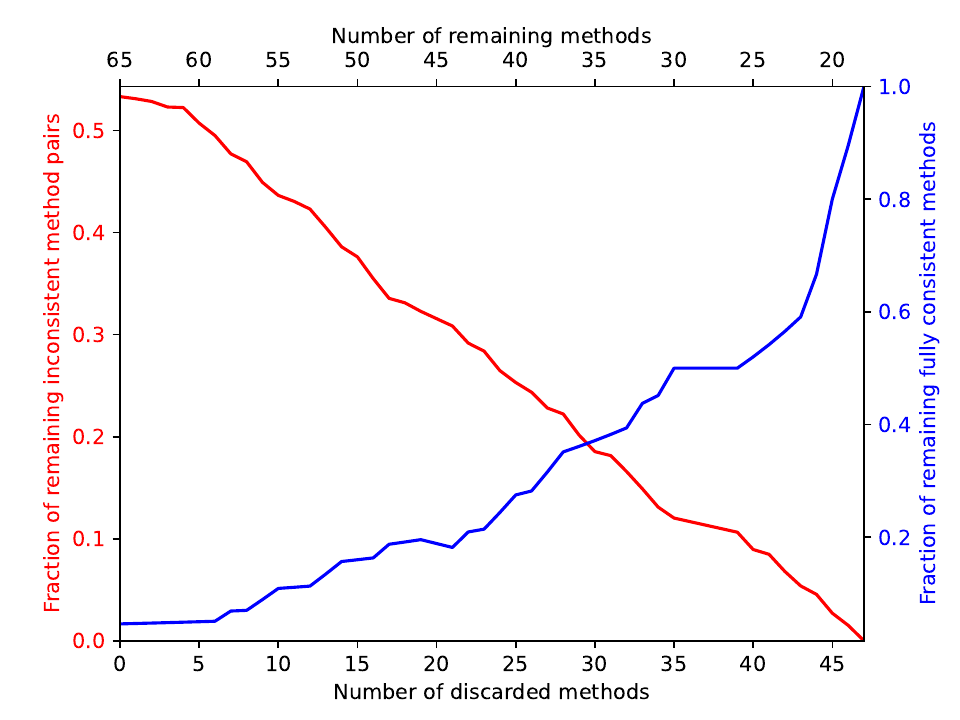}
    \caption{\textit{Red:} number of inconsistent method pairs (having $p < 0.05$) as a function of the number of discarded or remaining methods, assuming that discarded methods are responsible for inconsistency in all comparisons involving them. \textit{Blue:} number of fully consistent methods (having 100\% external consistency) as a function of the number of discarded methods. We also show the total number of methods remaining on the top axis. %\hd{Could you add a top axis which is number of methods remaining?}
    }
    \label{fig:consistency-evolution}
\end{figure}

% However, this by no means shows that these measurements of the expansion are subject to strong systematics. Collaborations like SH0ES use a smaller selected sample of distances \citep{riess2022comprehensive}.
For internal inconsistency, we show with an arrow whether the inconsistency indicates under- or overestimated uncertainties. Some indicators present underestimated uncertainties (\texttt{Horizontal Branch, AGN time lag, Type II Cepheids, G Lens, HII region diameter, GeV TeV ratio, and Mass Model}) while \texttt{Delta Scuti, Carbon Stars, Dwarf Galaxy Diameter, Ring Diameter, Faber-Jackson} and \texttt{Diameter} are the opposite. For the remainder ones, the cause of the inconsistency was not the inconsistent uncertainties. 
% If we constrained $H_0$ with overestimated uncertainties, we would derive a result with inflated error bars which would potentially bias results in the Hubble tension by underestimating the true tension.
Some distance indicators have external consistency results but no internal consistency $p$-value because they have at most one distance measurement to any given galaxy.
% However, this does not mean that other indicators can have shared measurements in the same galaxies.
One indicator (\texttt{GRB}) has an internal consistency result but no external consistency results because all of its measurements are to galaxies having no distances by any other method.

\begin{figure}
    \centering
    \includegraphics[width=1.05\linewidth]{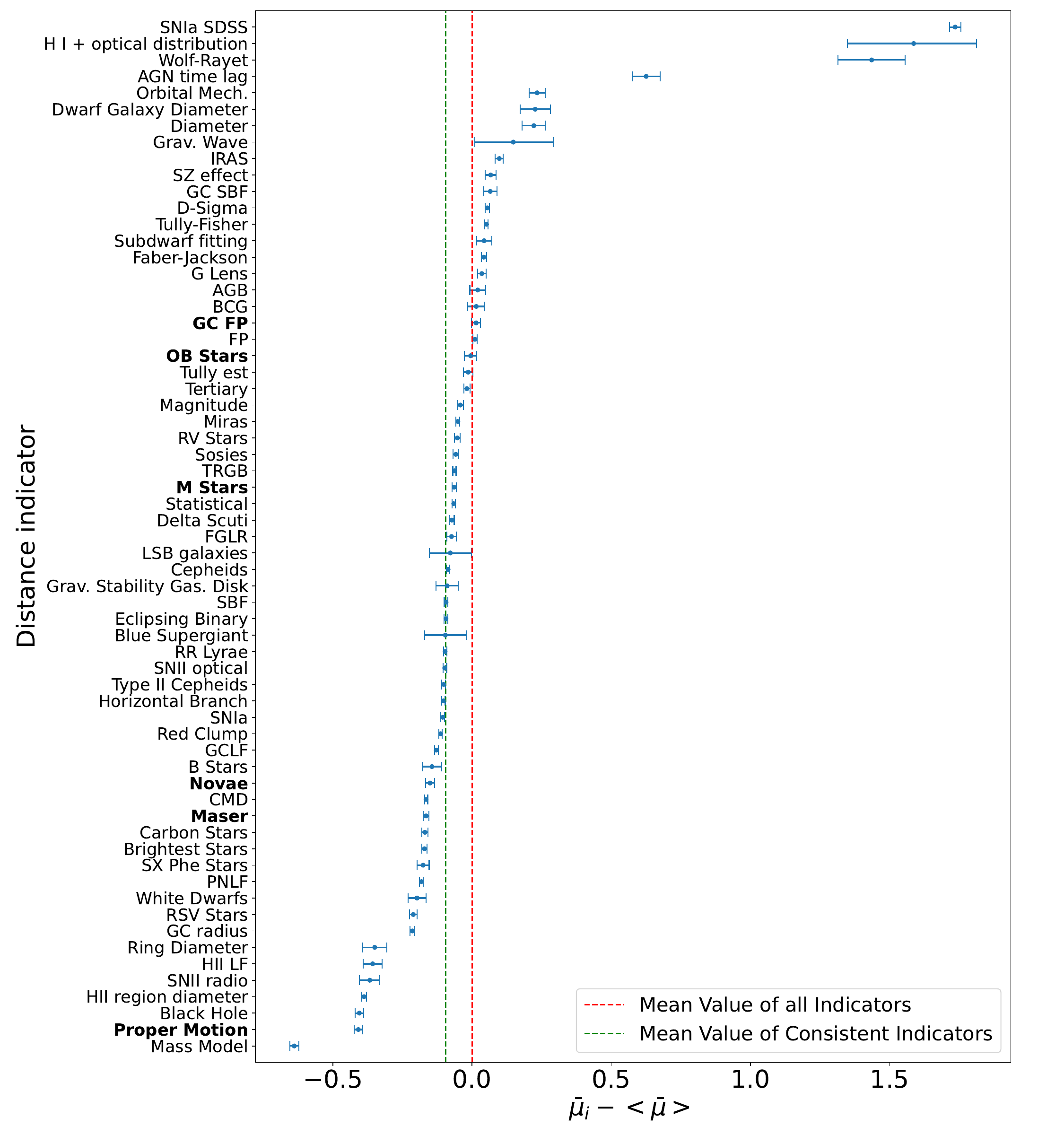}
    \caption{Mean values and $1\sigma$ C.L. uncertainties for the average distance modulus $\bar{\mu}$ returned by each indicator relative to the mean of all. Negative values correspond to distances underestimated on average relative to the mean of all indicators and vice versa.%\hd{Again bar notation or something} \hd{Enlarge what you can: the x label and numbers and ``Distance indicator''.}
    }
    \label{fig:whiskerPlot}
\end{figure}

In Fig.~\ref{fig:$p$-values} we summarize the results for all distance indicators. We show the $p$-values both for internal consistency along the diagonal and external consistency on the off-diagonal.
% results for the internal self-consistency (analysis of consistency of an indicator with itself) and the external-consistency (analysis of consistency of an indicator with all the other indicators). We show the results in a lower-triangle with the internal consistency results in the diagonal and the external consistency ones in the non-diagonal positions.
Cases where no consistency test is possible due to no overlap in measurements are shown as empty squares. Green indicates consistency ($p>0.05$), while $p$-values below 0.05 are shown by the colour-bar, with darker red indicating lower $p$. As we use 10,000 mocks the $p$-value is calculated in units of 0.0001; anything below this (i.e. no mock KLD as large as that of the data) is shown in dark red. The hatching indicates the cause of inconsistency. Over- and underestimated uncertainties are shown by forward and backward slanting diagonal lines respectively, while inconsistent means are shown by the circle pattern. Both types of hatching indicate that both problems are present to some degree, such that resolving either one produces consistency, and star hatching that both problems are present to a greater degree such that resolving both is required to remove the inconsistency. If neither shifting the means nor rescaling the uncertainties removes the inconsistency we label it ``essential'', indicated by no hatching.
% We also need to clarify that this kind of inconsistency is not mutually exclusive with inconsistent means or uncertainties. We can have for example a combination of outliers and inconsistent means or non-Gaussianity and overestimated uncertainties. Thus, these inconsistencies might have the previous problems studied but with some additional unrecognized inconsistencies. \\
% Finally, when the $p$-value is lower than 0.05, we use a colour-bar to determine the tone of the square. We used the \textit{Autumn} colour-bar from the \texttt{matplotlib.pyplot} python package \citep{Hunter:2007}. This gives a colour selection from red to yellow. In this way, we can visualize how far the $p$-value is from the threshold of 0.05. We used a logarithmic scale and set every $p$-value lower than $10^{-4}$ to red. When the $p$-value tends to 0.05, the colour is yellow, and an intermediate orange tone for intermediate values of the $p$-value.

Figure \ref{fig:$p$-values} contains all the information for the consistency of the distance indicators. The total number of consistent pairs is 427 (41.8\% of all pairs). 221 are inconsistent due to have uncertainties that are either too large or too small (mostly too small). Seven have inconsistent means, 109 are made consistent by matching either the means or the uncertainties, and five only by matching both. For problems that our tests cannot detect such as combinations of the previous ones along with e.g. outliers or non-Gaussian behaviour, we show no hatching. There are 163 such cases. Finally, 1994 pairs have no shared galaxies and hence no $p$-value. (Note that each pair is counted twice in Tables~\ref{tab:results-standard-candles} and~\ref{tab:results-standard-rulers}, so that the sums of the corresponding columns are twice the numbers quoted above.)
% Therefore, the analyses showing consistency are 427 out of 932, representing 41.8\% of all cases.
Combining \textit{a priori} consistent cases with those where we have identified the problem yields 769 cases, representing 82.5\% of the total. Hence, even though most distance indicators are internally inconsistent, close to half of the cross-analyses are consistent, while only around a fifth are inconsistent due to an undiagnosed problem.

\begin{figure}
    \centering
    \includegraphics[width=1.05\linewidth]{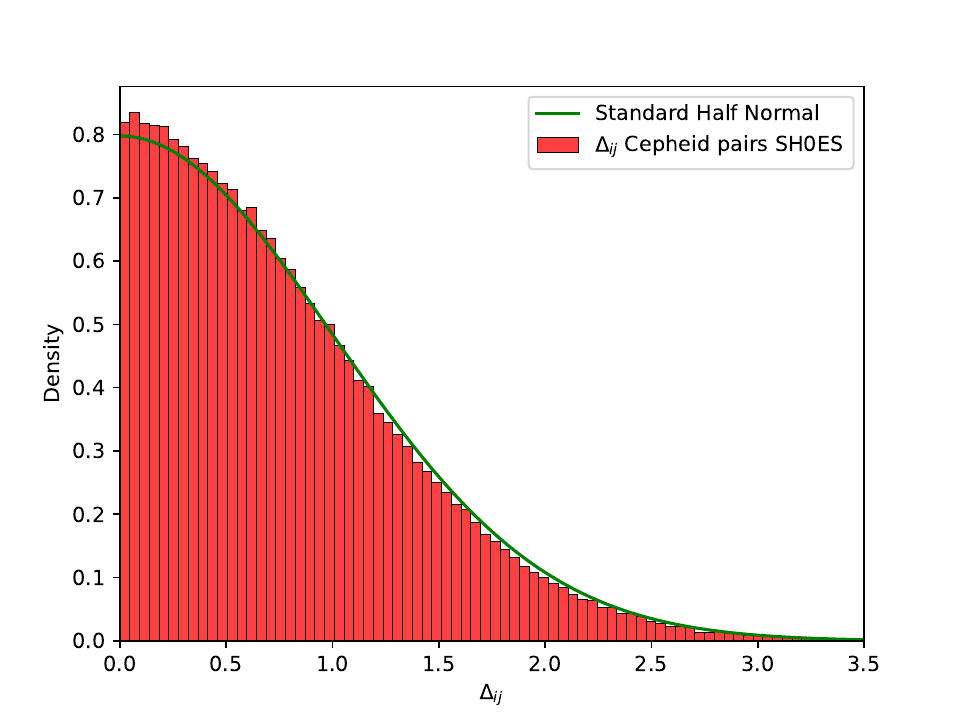}
    \includegraphics[width=1.05\linewidth]{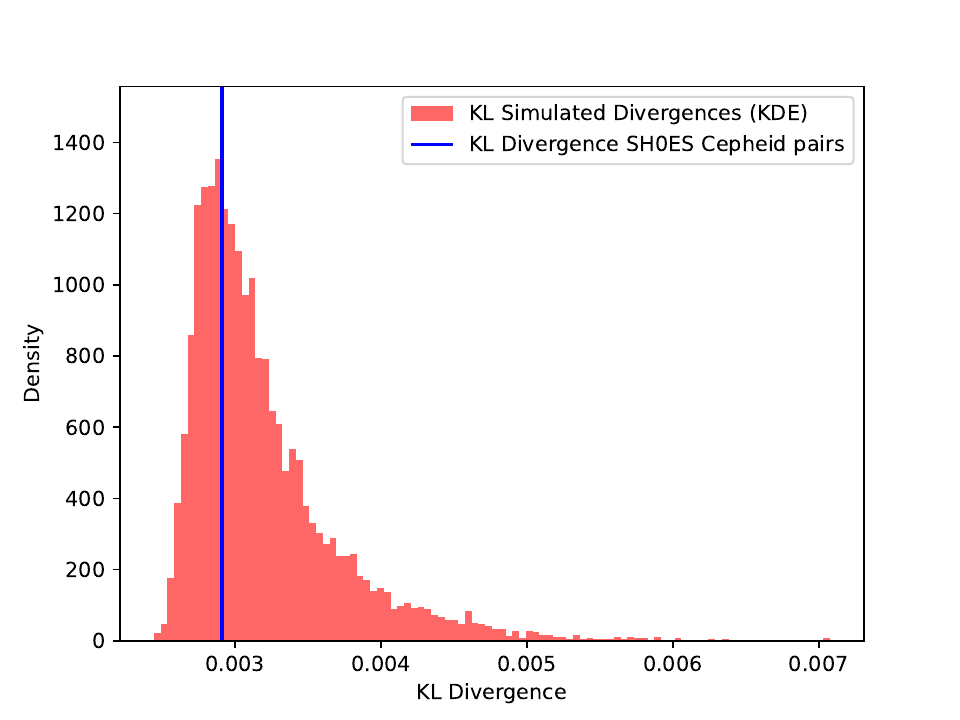}
    \caption{\textit{Top:} Normalized probability distribution function (pdf) of the difference in units of $\Delta_{ij}$ for the SH0ES 2022 catalogue. %\hd{$\Delta_{ij}$}.
    We show the theoretical pdf of the standard half-normal for comparison. %\hd{Truncate x-axis at 3.5.}
    \textit{Bottom:} Histogram for the mock KLD values generated under the null hypothesis of consistency compared to the KLD value of the real data.
    % We show the empirical result for SH0ES 2022 for a visual interpretation of the KL $p$-value.
    }
    \label{fig:SH0EShistogram}
\end{figure}
\begin{table}
    \centering
    \caption{As Table~\ref{tab:results-standard-candles} but for the SH0ES 2022 catalogue. In this case all distances use the Cepheid method so there is no notion of external consistency. N3982 and N0691 are shown in orange to indicate that they would fail a two-tailed consistency test (their $\Delta_{ij}$ values form a distribution more similar to a standard half-normal than would be expected from statistical noise under the assumption of consistency between all the measurements).}
    \begin{tabular}{ccc}
    \hline
       Name  & Measurement pairs & KL $p$-value \\ \hline
       Full dataset  & 345615 & \textcolor{OliveGreen}{0.6613} \\ \hline
       N4258 & 97903 & \textcolor{OliveGreen}{0.6896} \\
        LMC & 115921 & \textcolor{OliveGreen}{0.1286} \\ \hline
       N4038 & 406 & \textcolor{Orange}{0.9992} \\ 
N1015 & 153 & \textcolor{Orange}{0.9808} \\ 
N7678 & 120 & \textcolor{OliveGreen}{0.953} \\ 
N2525 & 2628 & \textcolor{OliveGreen}{0.933} \\ 
N1309 & 1378 & \textcolor{OliveGreen}{0.8774} \\ 
N7541 & 528 & \textcolor{OliveGreen}{0.7556} \\ 
U9391 & 528 & \textcolor{OliveGreen}{0.7356} \\ 
N4680 & 55 & \textcolor{OliveGreen}{0.661} \\ 
N0976 & 528 & \textcolor{OliveGreen}{0.6108} \\ 
N0691 & 378 & \textcolor{OliveGreen}{0.553} \\ 
N5584 & 13530 & \textcolor{OliveGreen}{0.5328} \\ 
N5728 & 190 & \textcolor{OliveGreen}{0.5194} \\ 
N1559 & 5995 & \textcolor{OliveGreen}{0.464} \\ 
N5861 & 820 & \textcolor{OliveGreen}{0.4522} \\ 
N3972 & 1326 & \textcolor{OliveGreen}{0.4354} \\ 
N5468 & 4278 & \textcolor{OliveGreen}{0.4062} \\ 
N4639 & 435 & \textcolor{OliveGreen}{0.3974} \\ 
N3254 & 1128 & \textcolor{OliveGreen}{0.3972} \\ 
N4536 & 780 & \textcolor{OliveGreen}{0.3938} \\ 
N3982 & 351 & \textcolor{OliveGreen}{0.2766} \\ 
N1365 & 1035 & \textcolor{OliveGreen}{0.2418} \\ 
N2608 & 231 & \textcolor{OliveGreen}{0.1936} \\ 
N3147 & 351 & \textcolor{OliveGreen}{0.1834} \\ 
M101 & 33411 & \textcolor{OliveGreen}{0.1768} \\ 
N7250 & 210 & \textcolor{OliveGreen}{0.1756} \\ 
N5917 & 91 & \textcolor{OliveGreen}{0.149} \\ 
N3447 & 5050 & \textcolor{OliveGreen}{0.1276} \\ 
N3021 & 120 & \textcolor{OliveGreen}{0.1174} \\ 
N1448 & 2628 & \textcolor{OliveGreen}{0.117} \\ 
Mrk 1337 & 105 & \textcolor{OliveGreen}{0.1142} \\ 
N3370 & 2628 & \textcolor{OliveGreen}{0.0778} \\ 
N2442 & 15576 & \textcolor{RedOrange}{0.042} (\textcolor{OliveGreen}{0.587}) $\uparrow$ \\ 
N3583 & 1431 & \textcolor{RedOrange}{0.0242} (\textcolor{OliveGreen}{0.2722}) $\uparrow$ \\ 
N5643 & 31375 & \textcolor{RedOrange}{0.0196} (\textcolor{OliveGreen}{0.851}) $\uparrow$ \\ 
N0105 & 28 & \textcolor{RedOrange}{0.0166} (\textcolor{OliveGreen}{0.4116}) $\downarrow$ \\ 
N7329 & 465 & \textcolor{RedOrange}{0.0154} (\textcolor{OliveGreen}{0.6016}) $\downarrow$ \\ 
M31 & 1485 & \textcolor{RedOrange}{0.015} (\textcolor{OliveGreen}{0.643}) $\downarrow$ \\ 
N4424 & 36 & \textcolor{RedOrange}{0.0004} (\textcolor{OliveGreen}{0.1314}) $\downarrow$ \\ \hline
    \end{tabular}
    \label{tab:SH0ESResults}
\end{table}

In Fig.~\ref{fig:consistency-evolution}, we present the percentage of inconsistent method pairs and of fully externally consistent methods. This is shown as a function of the number of discarded or remaining methods (see Sec.~\ref{sec:ins_inc}), where it is assumed that each discarded method is responsible for inconsistency in any pair in which it is present. We continue discarding methods until we obtain the maximum set of fully externally consistent methods. We find this set to contain the following 18 methods: \texttt{AGN time lag, PNLF, Sosies, M Stars, LSB
galaxies, Proper Motion, Novae, SX Phe Stars,
Maser, Dwarf Galaxy Diameter, GC FP,
Subdwarf fitting, OB Stars, Orbital Mech. , B
Stars, AGB, BL Lac Luminosity, GeV TeV ratio}. It is interesting to compare this with the set of 12 self-consistent methods: \texttt{BL Lac Luminosity, RSV Stars, OB Stars, Novae, Miras, RV Stars, Statistical, Brightest Stars, M Stars, Proper Motion, Maser, GC FP}. The intersection of these sets gives us a set of 7 methods simultaneously achieving perfect internal and external consistency, which we highlight in bold throughout the paper: \textbf{M Stars, Proper Motion, Novae, Maser, GC FP, OB Stars, BL Lac Luminosity}. If we showed Fig.~\ref{fig:$p$-values} with only these indicators, all the non-empty squares would be green.

The maximum set of fully externally consistent methods is not the same as the set of self-consistent methods. The set of methods being fully externally consistent but self-inconsistent contains 11 methods: \texttt{AGN time lag, PNLF, Sosies, LSB Galaxies, SX Phe Stars, Dwarf Galaxy Diameter, Subdwarf fitting, Orbital Mech., B Stars, AGB, GeV TeV ratio}. On the other hand, the set of methods being self-consistent but not fully externally consistent contains 5 methods: \texttt{RSV Stars, Miras, RV Stars, Statistical, Brightest Stars}. It is interesting to see that being fully externally consistent does not guarantee being internally consistent. This can be caused by a consistent subset of an indicator agreeing with other distance indicators or by both indicators being inconsistent in a way that cancels out when they are compared.
% and seeming consistent together.

In Fig.~\ref{fig:whiskerPlot}, we present a whisker plot for the mean value and minimum $1\sigma$ C.L. uncertainties for the constraints on $\bar{\mu}_i$%\hd{again, bar or something}
, a measure of the overall distance modulus returned by each method relative to the average of all of them. We present them in order for visual clarity.
% \hd{Are we using the same order elsewhere? I think we should.}.
We discard two indicators (\texttt{BL Lac Luminosity} and \texttt{GeV TeV ratio}) as they both had only one constraint equation, which linked them (i.e. they each overlapped with only one other method in one galaxy, and that method for one of this pair was the other). This causes a perfect degeneracy between their $\bar{\mu}$ values, making the marginalised posteriors unbounded. %\hd{Describe the results a bit in terms of what good and bad methods are as indicated by this, how many methods give results consistent with the mean here, or a certain number of sigma discrepant.} \\

Fig.~\ref{fig:whiskerPlot} shows the methods that tend to overestimate and underestimate distance modulus in average. \texttt{Mass model} is the indicator with the strongest underestimation having a difference $\Delta \bar{\mu} \equiv \bar{\mu}_i - \left<\bar{\mu} \right> \approx -0.6$ which corresponds to an underestimation of the luminosity distance by $\sim$20\%. \textbf{Proper Motion}, \texttt{Black Hole, HII region diameter, SNII radio, HII LF, and Ring Diameter} have a difference of lower then -0.25 implying underestimation of the distance relative to the mean by more than 10\%. Three indicators (\texttt{SNIa SDSS, H I + Optical distribution} and \texttt{Wolf-Rayet}) have a difference of around $1.5$ and higher, representing an overestimation of distance relative to the mean by a factor of around 2 or more.
% . This represents an overestimation of the overall distance higher than 95\%.
% This is a sign of very strong systematics and problems as they almost double the distances.
If we look for these indicators in figure \ref{fig:$p$-values}, we can see that they are not consistent in any of the analyses. This is expected since they appear to overestimate distances by a huge margin. \texttt{AGN time lag} shows a difference higher than 0.5 implying an overestimation of distances higher than 25\%. %\texttt{Dwarf Galaxy Diameter, Diameter and Grav. Wave} have a difference higher than 0.25 (for \texttt{Grav. Wave}, this is the case just if we consider the upper error bars as the lower ones are close to the mean) which imply an overestimation of distance higher than 12\%. 
Out of the 7 indicators achieving full external consistency and internal consistency, only one seems to be significantly misestimating distances relative to the mean: \textbf{Proper Motion}. %\textbf{Maser} is close to this indicator but has an underestimation of $\sim$12\%. 
The remainder of the fully consistent indicators return values close to the mean (\texttt{BL Lac Luminosity} has no constraint due to the perfect degeneracy with \texttt{GeV TeV ratio}). Note that of course the mean may be biased by inconsistent methods, so returning distances compatible with the mean is not a reliable indicator of a method's quality.

\subsection{SH0ES}\label{sec:results_shoes}

We now move to the analysis of the independent SH0ES 2022 catalogue. Fig.~\ref{fig:SH0EShistogram} presents two histograms for this catalogue. The upper panel shows the normalized pdf %\hd{this is sometimes capitalised and sometimes not}
of the differences between the distance moduli ($\Delta_{ij}$) and the comparison to the standard half-normal. The lower panel shows the histogram of KLD values over the 10000 mock datasets compared to the value in the real data (vertical blue line).
% We show the empirical KDE KL divergence for comparison and for a visual interpretation of the $p$-value.
The full results are shown in Table \ref{tab:SH0ESResults}, which (except for the first row describing the full catalogue) contains the galaxy name, number of measurement pairs and KLD $p$-value. The first two galaxies (N4258 and LMC) are in the first rung of the distance ladder and are used to calibrate the Cepheid PL relation by means of geometric distances to those galaxies.  %\hd{Give a short description of the results, including the two that are too consistent for their own good.}\\

Fig.~\ref{fig:SH0EShistogram} and the top row of Table~\ref{tab:SH0ESResults} indicate that the SH0ES 2022 catalogue agrees well with the null hypothesis of statistical consistency. The two galaxies used for the first rung that calibrates the PL relation are also consistent with the null hypothesis. It is particularly important for these to be consistent, as inconsistency there would propagate into all the other distances and hence the constraint on e.g. $H_0$. From the 38 galaxies in the second rung (M31 does not host a SNIa but it is included in the SH0ES catalogue), 31 are consistent while the remaining 7 are inconsistent due to underestimated uncertainties. Note that most of the $p$-values only narrowly pass the threshold for rejection of the null hypothesis of consistency; had we take a tighter 0.01 threshold for the $p$-value, only one galaxy would have been considered inconsistent (N4424). This contrasts with NED-D where the great majority of inconsistent cases have $p<0.01$.
% The inconsistency \purp{has} in most cases \purp{$p> 0.01$, which is higher than most NED-D analyses};  \purp{with the null hypothesis}.
Most of the inconsistent galaxies have $<$2000 pairs (apart from N2442 and N5643) and hence play a relatively small role in the consistency of the whole catalogue and the calibration of $M^0_{B}$ and hence $H_0$. Four of the seven inconsistent galaxies have underestimated uncertainties (N0105, N7329, M31 and N4424) whereas N2442, N3583 and N5643 have overestimated uncertainties. Interestingly, the inconsistent galaxies having more than 15000 measurement pairs have overestimated uncertainties, which has the effect of overestimating the uncertainty on $H_0$. We highlight two galaxies in lighter orange (N4038 and N1015) that would be inconsistent under a two-tailed test due to having $p>0.975$.
% In this test, we need to consider galaxies with lower $p$-value than 0.025 and higher than 0.975 as inconsistent and not just lower than 0.05.
This is due to the KLD being surprisingly low relative to the mock data distribution, indicating that the data conforms more closely to a standard half-normal than would be expected under the assumption of consistency given the uncertainties.
% The cause of the inconsistency in the upper tail is due to the KL divergence being surprisingly low, way lower as we expect from a random distribution. For the two tail test, we would have two more inconsistent galaxies but \textbf{N1559} would not be inconsistent and so the total number of inconsistent galaxies would be 8.

% In general, the SH0ES 2022 shows reasonable consistency with the null hypothesis and hence our analysis does not find evidence for systematics that could impact the best estimate or uncertainty of the Hubble constant.
% This implies that we found no evidence for the Hubble-Lemaitre tension coming from inconsistent means and uncertainties in the SH0ES 2022 catalogue.
% It is possible however that this consistency may be to some extend enforced by analysis choices made in the SH0ES pipeline, for example rescaling uncertainties to produce a reduced $\chi^2$ of 1~\citep{riess2022comprehensive}.

To study the effect of the inconsistent galaxies on the constraints on $H_0$ in SH0ES, we repeat the MCMC inference described in Sec.~\ref{sec:data_shoes} but removing the seven inconsistent galaxies. These are all second-rung or non-SNIa host galaxies.
% Thus, by excluding these data, the constraints on $M^0_B$ change and thus on $H_0$.
The new constraint on the Hubble constant is $5 \log_{10}H_0 = 9.319 \, \pm \, 0.032$, or $H_0 = 73.08^{+1.07}_{-1.08}$ km/s/Mpc in comparison to the baseline result with all galaxies of $5 \log_{10}H_0 = 9.318 \, \pm \, 0.031$, or $H_0 = 73.04 \, \pm \, 1.01$ km/s/Mpc. Both are at a $5\sigma$ C.L. tension with Planck 2018.
This is a negligible change from the case with all galaxies included, showing that the Hubble tension does not result even in part from these galaxies with inconsistent Cepheid distances.
% The difference of this reanalysis with respect to Planck 2018 flat $\Lambda$CDM \citep{aghanim2020planck} in units of $\Delta 5 \log_{10} H_0$ is $0.179 \, \pm \, 0.035$, giving a difference of $5\sigma$ C.L. The constraints on the Hubble constant change slightly as the mean and error bars increase. However, the tension with Planck 2018 remains at $5\sigma$ C.L. This shows that even by only including the fully consistent galaxies, the tension remains the same.}

\section{Discussion}
\label{sec:disc}

%\hd{Most of the text up to sec 5.1 belongs in the Results section. Anything that is a description of the results, e.g. observations from a plot, belongs there. This section is for more general discussion, e.g. that use the results of multiple plots or that connect to broader issues like the Hubble tension. Can you try to make this division? What stays in this section should go into one of the subsections below: there shouldn't be anything between 5 and 5.1.}

\subsection{Implications of the results}

Our main result is that many of the distance measurements in NED-D are subject to significant problems, including ``unknown unknowns'' such as unaccounted-for systematics and outliers as well as the ``known unknowns'' of misestimated uncertainties and inconsistent central distance values between indicators. 
This makes it challenging currently to use for inference of cosmological parameters; the only previous attempt to do so yielded a 21\% precision estimate of $H_0$ ($70 \pm 15$ km/s/Mpc) from an ensemble of estimates and galaxies comparable to that studied here~\citep{nedd_mean}. This is likely sufficiently low precision not to be significantly affected by the inconsistencies we identify here.
% This motivates the endeavours of collaborations like SH0ES to construct smaller and more carefully curated, and hence more consistent, extragalactic distances datasets. 
Despite our results, NED-D provides an excellent starting point to build a robust set of distance ladder measurements for cosmological inference. This would be similar to the approach of collaborations like SH0ES that construct and employ smaller, more carefully curated---and hence more self-consistent---extragalactic distance datasets. 

Indeed, we find the Cepheids in the SH0ES 2022 catalogue not to exhibit significant problems. This supports the findings of~\citet{riess2022comprehensive} that the dispersion in the Cepheid measurements agrees with the fiducial uncertainties without any unexplained variance. However, it is important to note that our test is not sensitive to constant shifts in the distance modulus across all Cepheids in a given galaxy. This would leave the distance modulus difference between the Cepheids, $\Delta_{ij}$, unchanged. This effect could in principle be present if the Wesenheit relative magnitude $m^W_H$ or the fiducial Wesenheit absolute magnitude $M^W_{H,1}$ have an unaccounted-for constant systematic error. Such a global shift to the distance modulus to each Cepheid could alter the Hubble diagram arbitrarily when combined with SNIa.
% Thus, although our test shows no evidence for unaccounted systematics coming from inconsistent means or uncertainties, it cannot rule out this scenario.}
%This analysis is complementary to others studying possible systematics and model misspecification in the Cepheid--SNIa determination of $H_0$ \citep{Becker, rigault2015confirmation, rigault2020strong, kushnir2024cepheid, sharon2024reassessing, riess2024jwst, uddin2024carnegie, riess2024jwst2}.

There is a small set of methods that exhibit both internal and external consistency (\textbf{M-stars luminosity, Novae, Masers, Globular Cluster Fundamental Plane, O- and B-type Supergiants, BL Lac Luminosity} and \textbf{Proper Motion}). However all of these have a small number of measurement pairs (no more than 127), and hence would not afford good constraints on the distance ladder---or cosmological parameters derived therefrom---in their own right. A worrying observation in this regard is that it is harder to acquire sufficient evidence to reject the null hypothesis when the dataset is smaller, suggesting that these methods may also become classified as inconsistent as more data on them is collected.

It is important to emphasize that statistically consistency is a necessary but not sufficient condition for a method to be useful in the distance ladder. Another necessary condition is to have a sufficient number of measurements to bring significant constraining power. For this reason, the seven methods in our fully consistent set are inferior to others having many more measurements like \texttt{Cepheids, CMD, RR Lyrae, TRGB, Tully-Fisher} and others. Future work must therefore either focus on statistically consistent subsets of these methods, or develop ways of measuring distances using the methods we identifying as having naturally higher consistency to more galaxies.
% The objective for future work is thus to build a reliable subset of distances using these methods.
An interesting indicator in this regard is \texttt{Faber-Jackson}, which has a very high number of measurement pairs (1947) and a p-value of 0.0101, near the threshold for consistency. We see from Table~\ref{tab:results-standard-candles} that the inconsistency is due to overestimated uncertainties, probably the least pernicious cause. Another interesting case is \texttt{CMD}, which is self-consistent when only taking the measurements standardised through $H_0$ or $\mu_{LMC}$ (Appendix~\ref{sec:app}) Moreover, this indicator has 3454 measurement pairs, almost double \texttt{Faber-Jackson} when restricting to standardisable measurements. It therefore shows promise as a useful method for constraining the distance ladder in the future.
% case. This shows that when we take suitable subsets of the data, we can achieve consistency and a high enough number of pairs, two necessary conditions to reconstruct the distance ladder,}

Containing as it does measurements collected over 40 years, it is perhaps unsurprising that the overall consistency of NED-D is low. To investigate this further, we try splitting some of the better-known methods with many measurements (\texttt{Cepheids, TRGB, SNIa, RR Lyrae, CMD} and \texttt{Tully-Fisher}) into three bins of measurement year with roughly equal numbers of measurements in each bin and calculating the $p$-value for self-consistency within each bin. In this way, we can see if there are signs of an improving consistency over time. However, we find that for all indicators and bins the $p$-value remains, with the sole exception of \texttt{RR Lyrae} which has a $p$-value of 0.0008 for the intermediate bin. This indicates that the inconsistencies are not solely due to poorer measurements in the past or discrepancies between analysis choices in the past versus more recently.
% Thus, we see no improving consistency over time and other approaches need to be taken to select a consistent subset of measurements capable of reconstructing the distance ladder.

Some of the inconsistency derives from non-independent measurements. For example, NED-D includes three TRGB distances to N6822 from \citet{fusco2012distance}, yet two of these differ only in the TRGB calibration, not the measurements themselves, and the third is an average of these two. This is simple a limitation with the way the literature was trawled to compile the catalogue: the abstract of the paper itself presents just one independent measurement. Another example comes from \citet{rich2014new}, where the distances to 39 Cepheids in the galaxy NGC 6822 are provided. However, six of them were excluded from the analysis in the paper due to problems like phase shifting, crowding and deviations from the PL relation. Yet all 39 measurements are included in NED-D, potentially contributing significantly to  inconsistency.

By excluding these problematic and non-independent measurements, we would get more information on the possibility of reducing the inconsistency across NED-D.
% between the distances given by the indicators.
One way to do this might be to automate a search over the literature using an Artificial Intelligence Large Language Model (LLM) as opposed to the human searchers employed in the construction of NED-D. This could allow a pre-cleaning of the data capable of removing unreliable cases.
% ModelOne way to do this would be to go over the original references and remove the problematic measurements from the data. This can be automated with a Large Language Model-based process that identifies the unreliable cases.
The LLM could also potentially be used to homogenise the calibrations of the methods by identifying the anchor used in a given paper, and its assumed distance, and then rescaling the measurements appropriately.
If the inconsistencies could be reduced sufficiently, enormously more measurements---from a great deal more distance indicators---could be used to calibrate the distance ladder for cosmological inference and other applications.
% By making a pre-cleaning of the data, the inconsistencies might be reduced enough to enable constraints of $H_0$ with some of the NED-D distance indicators.

Our results could potentially help explain the Hubble tension. For example, Fig.~\ref{fig:whiskerPlot} indicates that Cepheid distances are lower than the mean of all methods. If distance were underestimated in SNIa host galaxies, then, by $H_0 \approx cz/D$, the Hubble constant would be overestimated. TRGB likewise produce distances below the mean. Note however that the mean of all methods (most of which are inconsistent) is not a reliable indicator of the truth, and this simple accounting neglects the possibility of \emph{relative} differences between the calibrator and cosmological first-rung samples.

To close this subsection we briefly describe the physics of the methods that we find to be most consistent, achieving both internal and external consistency. \textbf{M-stars luminosity} uses a relationship between the absolute magnitude of red dwarf stars and the temperature-independent spectral index. However, it requires low-resolution spectroscopy and thus it can only be applied in nearby galaxies, like the LMC \citep{schmidt1998luminosity}. The \textbf{Novae} standard candle comes from a relation between the absolute magnitude at the maximum and the rate of decline in the flux known as the MMRD relation \citep{della1994nova}.
% The determination of the absolute magnitude of novae using this relation gives $M_V \sim -8.77 \, \pm \, 0.3$ \citep{ferrarese1996discovery}.
For \textbf{O- and B-type Supergiants} stars, there exists a relation between their absolute magnitude and their spectral type and luminosity class. This method was notably applied to 30 Doradus nebula in the LMC \citep{walborn1997spectral}. \textbf{BL Lac Object Luminosity} is a method based on calibrating the absolute magnitude of BL Lacertae objects. It was found that the dispersion of this is small, although standardising it robustly does require a K-correction and evolution correction \citep{sbarufatti2005imaging}.

The \textbf{Maser} standard ruler is based on the dynamics of a water maser in an accretion disk surrounding a supermassive black hole. Using Very Long Baseline Interferometry (VLBI) it is possible to determine individual maser spots,
% which help determine the position,
the velocity of the masers and the shape of the disk. The physical size can then be deduced from Kepler's laws and the angular size, affording a determination of the angular diameter distance \citep{humphreys2004improved, reid2019improved}. The \textbf{Globular Cluster Fundamental Plane (GCFP)} relates velocity dispersions, radii, and mean surface brightness of globular clusters to their absolute magnitude. This has been used to determine the distance to M31 \citep{strader2009mass}. Finally, the \textbf{Proper Motion} method compares a galaxies' apparent motion on the plane of the sky to its absolute motion calculated dynamically. This is also only applicable locally where proper motions are measurable (e.g.~\citealt{lepine2011first}).

%\hd{To add: something about the physics of the (best) methods}
% after discussing w/ Claudia}
% \jan{To be done}

\subsection{Comparison with the literature}

Our NED-D analysis builds upon \citet{Singh}, where possible evidence for systematic underestimation of distance uncertainties was found. This could imply strong systematics on the late-time measurements of $H_0$, for example \citep{cchp_trgb, riess2022comprehensive, anand2022comparing, scolnic2023cats, Riess_review}, which might overestimate the tension with \textit{Planck}. %However, as mentioned before, \citet{Singh} uses a non-symmetric expression for the discrepancy between two measurements that does not include the uncertainty of both measurements. This is in contrast to the Mahalanobis distance (Eq.~\ref{eq:differenceSigma}) which takes both of them into account. The non-symmetric expression gives higher sigma differences as one of the uncertainties is neglected, artificially increasing discrepancies. Even so, i
Despite using a different metric to quantify tension, and a considerably more sophisticated statistical framework, we agree with ~\citet{Singh} that most NED-D methods are inconsistent.

The NED-D catalogue has been used in various ways. For example, \citet{cappellari2011atlas3d} estimated uncertainties in the distances to galaxies in the ATLAS3D sample estimated from a Virgocentric infall peculiar velocity model by correlating them against the direct measurements in NED-D. This produced a sizeable 27\% distance uncertainty for 692 galaxies with at least one NED-D distance measurement. \citet{ohlson202350} used NED-D along with the Local Volume Galaxy catalogue \citep{karachentsev2013updated} to create a catalogue of galaxies within 50 Mpc with known distances, to study for example their nuclear X-ray activity. 
% They used 285 galaxies in common and correlated the median of the $D_{NED} \, H_0$ from NED-D against their cosmic velocities \citep{cappellari2011atlas3d}.

NED-D has also been used to constrain fifth forces induced by modified gravity.
% catalogue can also be used to constrain $\Delta G/G_N$, the change of the effective Newton's constant in modified theories of gravity. For example, in
\citet{desmond2019local} used the measurements of 51 galaxies with Cepheids and TRGBs from NED-D to set a bound on $\Delta G$, an effective change to Newton's constant due to a long-range fifth force. This test assessed only \emph{relative} consistency between Cepheid and TRGB distances as a function of $\Delta G$, although we see from Fig.~\ref{fig:$p$-values} that these indicators are essentially inconsistent in NED-D. The consistency may be worse under modified gravity, but it is not good under General Relativity. Similarly, \citet{desmond2021five} used the TRGB distance to the LMC from NED-D as a constraint in inferring the gravitational constant in this dwarf galaxy from the dynamics of Cepheids in detached eclipsing binary systems. Each of these studies may be impacted by the inconsistencies we unearth here, and modified gravity constraints based on direct distance comparisons is a topic we will revisit in future work.

Our SH0ES analysis is complementary to others studying possible systematics and model misspecification in the Cepheid--SNIa determination of $H_0$. For example, the error bars on $H_0$ are impacted by outliers present in the data \citep{Becker}. This has been addressed in more recent work by removing the Cepheids that differ from $3.3\sigma$ in the global fit of the PL relation, which represent 1.2\% of the catalogue \citep{riess2022comprehensive}. Another potential systematic is an environment-dependence on SNIa absolute magnitude, in the sense that SNIa in locally star-forming environments are dimmer than those in passive environments \citep{rigault2015confirmation, rigault2020strong}. Another consideration is a possible difference between Cepheid observations and data reduction in the more nearby anchor galaxies (MW or LMC) relative to farther away ones like N4258. This can be studied by using only N4258 as the anchor~\citep{kushnir2024cepheid}. This possibility seems less likely in light of our tests assessing consistency in individual galaxies, in particular the finding that both the LMC and N4258 have internally consistent Cepheid distances. The inferred value of $H_0$ can also be passband-dependent, with \citep{uddin2024carnegie} suggesting lower values in the B-band relative to the H-band used in SH0ES. Finally, the James Webb Space Telescope (JWST) has provided new insights on the possible systematics in the distance ladder. In particular, it has been claimed that JWST validates the HST distance measurements in N4258 and five SNIa hosts, rejecting at high significance the possibility of unrecognized crowding of Cepheid photometry growing with distance \citep{riess2024jwst}. Indeed, some early samples of JWST Cepheid distances calibrated with only N4258 agree at with their HST counterparts \citep{trgb_ceph_comp, riess2024jwst2}.

Nevertheless there is a plethora of possible problems with the Cepheid distance ladder~\citep{problem_1,problem_5,problem_2,problem_3,problem_4}, as well as results from the Chicago-Carnegie Hubble Program (CCHP) using new data from the James Webb Space Telescope (JWST) indicating that $H_0\approx70$ km/s/Mpc with an insignificant tension with \textit{Planck}~\citep{cchp_jwst, cchp_jwst_2}. This agrees with earlier CCHP results based on TRGB~\citep{cchp_trgb}, and includes distances measured by the novel J-region Asymptotic Giant Branch (JAGB) method which boasts a number of advantages over Cepheids and TRGB~\citep{jagb_1,jagb_2}.

\section{Conclusions}
\label{sec:conc}

We study the consistency of 66 different methods for estimating redshift-independent distances from the NASA/IPAC Extragalactic Database of Distances (NED-D;~\citealt{steer2016redshift}) and SH0ES 2022 \citep{riess2022comprehensive} catalogues.
% NED-D includes redshift-independent determinations of extragalactic distances for 76 different distance indicators.
This is achieved by selecting galaxies that have at least two distance measurements by a given method (``internal consistency'') or at least one measurement by each of two different methods (``external consistency''). We use the Mahalanobis distance to compute the distance differences in units of sigma, forming a distribution for each consistency test. If the distance estimates are consistent (scattered from a common value by their quoted uncertainties), this distribution should form a standard half-normal. To test if this is the case, we compute the Kullback–Leibler (KL) divergence of the empirical distribution and of 10,000 realizations of random samples under the null hypothesis of consistency and according to the measurements' uncertainties. We compute the $p$-value as the fraction of realisations with a KL divergence higher than the value in the real data, adopting a threshold of 0.05 to discard the null hypothesis of consistency. This indicates significant evidence for under- or overestimated uncertainties, inconsistent means, statistical outliers, non-Gaussian uncertainties and/or systematic errors. When $p > 0.05$, there is not enough evidence to discard the null hypothesis, which we denote by ``consistent.'' We perform a similar analysis for the SH0ES 2022 catalogue.

For NED-D we find that only
% 66 of the distance indicators have at least one galaxy with either multiple measurements by that method or a measurement by another method, such that a consistency analysis is possible. We find that only
12 distance indicators achieve internal consistency. The maximal set of perfectly \emph{externally} consistent methods has size 18 (i.e. all methods in this set are consistent with each other).
% \hd{Is this actually what we're calculating, the largest set of mutually self-consistent methods?} \jan{Externally consistent, that's why I added that word now}.
The intersection of both sets gives the seven most reliable methods, which are both internally and externally consistent; these are \textbf{M Stars luminosity, Proper Motion, Novae, Maser, Globular Cluster Fundamental Plane, O- and B-type Supergiants} and \textbf{BL Lac Object Luminosity}. However, their number of measurements with these methods is low, generally $\mathcal{O}(10)$. This makes them unsuitable for a reconstruction of the distance ladder as they currently stand.
% Thus, statistical consistency is a necessary but not sufficient condition for this purpose.
A necessary condition besides statistical consistency for usefulness in the distance ladder is having a high number of measurements, and thus a better approach than focusing on these methods may be to build a consistent subset of distances using the indicators that provide many measurements. This paper gives a starting point for such constructions.

We also calculate the average measured distance modulus of each indicator across all galaxies relative to the mean of all indicators. We find that from the best set of indicators, \textbf{Proper Motion} is the only one with a considerable discrepancy in average returned distance (17.1\% underestimation) relative to the mean.
% This indicator gives a luminosity distance underestimation higher than 10\% with respect to the mean whereas the remainder ones have a performance close to the mean.
Outside the set of best indicators, some overestimate the luminosity distances by more than 95\%; these are \texttt{SNIa SDSS, H I + optical distribution} and \texttt{Wolf-Rayet}. \texttt{AGN time lag} shows an overestimation higher than 25\% and \texttt{Mass model} an underestimation higher than 20\%. This indicates that some distance methods in NED-D are strongly affected by systematics and thus unsuitable for cosmological studies. We perform 932 analyses in total, including both internal and external consistency, of which 427 are consistent (41.8\% of the total and 12 of them are from internally consistency analyses and the remainder externally consistency analyses). We repeat the analyses for the inconsistency cases by shifting the means and rescaling the uncertainties and find that 342 cases are made consistent by one or more of these modifications.
% Thus, for them we find the cause of the inconsistency. So, we achieve an updated consistency for 610 out of the 932 cases (representing 65.45\% of the total). This shows that even though most indicators are internally inconsistent, we achieve consistency or identified the problem causing the inconsistency for almost two thirds of the analyses.} %\hd{Maybe also mention some other results from fig 3, like what the most extreme methods are, how many are inconsistent with the mean at $>5\sigma$, etc.} \hd{Also say something about what tensions can be removed by shifting the means and/or uncertainties.}\\
In the remaining $\sim$1/5 of cases the inconsistency is not due only to incorrect central values or uncertainties, and we cannot identify the cause.

The SH0ES 2022 catalogue exhibits better behaviour. The KL divergence of the full catalogue yields a $p$-value of 0.66, in agreement with the null hypothesis of consistency. There is thus no evidence for hidden systematics in the Cepheid distances of this catalogue.
% and thus no evidence of inconsistent means or uncertainties as the cause of the Hubble-Lemaitre tension.
We repeat this analysis for each of the 40 galaxies in the catalogue separately. The two galaxies in the first rung (N4258 and LMC) that calibrate the PL relation have consistent $p$-values as well (0.69 and 0.13). This is crucial because an inconsistency in the first rung would propagate to the second and third rungs, potentially biasing the entire distance ladder.
% result of the Hubble constant $H_0$.
From the second-rung galaxies, we find that 7 out of the 38 present an inconsistency ($p < 0.05$) in a one-tailed test, and 8 in a two-tailed test. These galaxies however play a relatively minor role in the
% Even so, by taking the overall behaviour of the catalogue, these galaxies play a small role in the consistency and in the
calibration of the SNIa absolute magnitude for which the catalogue was designed. We show this by repeating the MCMC excluding these inconsistent galaxies, finding very similar $H_0$.
% an almost identical constraint on $H_0$.
% We derived $5\log_{10} H_0 = 9.319 \, \pm \, 0.032$, in $5\sigma$ C.L. tension with flat-$\Lambda$CDM. This proves that the removal of inconsistent distances keeps the tension at the same level and thus they are not a cause of it.
% \hd{What is the evidence for this?} \jan{Our test excludying these galaxies will show if this is true or not}.

In summary, the NED-D distance indicator measurements are subject to problems such as systematics, outliers, non-Gaussian behaviour and misestimated uncertainties. This is perhaps unsurprising given the heterogeneity of the measurements included, going back $\sim$45 years. Thus the full catalogue is not suitable for construction of a robust distance ladder. Even so, several papers have used the NED-D data for studies in cosmology \citep{freedman2019carnegie, chaparro2019predicting, nedd_mean, zaninetti2023classical}. By selecting a reliable subset of distances, it is however possible to determine galaxies' distances accurately. Some of the inconsistency comes from non-independent and unreliable distance measurements that were impossible to exclude in the search procedure used to construct the catalogue. By removing these problematic data, the overall consistency might be considerably enhanced, enabling cosmological analysis; we will report elsewhere on our attempts in this direction. The more homogeneous and quality-controlled SH0ES 2022 catalogue, however, has no significant evident problems.
The framework we develop may be simply applied to any catalogue of distances (or other quantities) to assess their consistency;
% We hope the framework devised here
we therefore hope that it will contribute to the future construction of a multiply cross-checked and fully robust distance ladder employing a wide range of methods.
% Finally, we found that the SH0ES 2022 catalogue \citep{riess2022comprehensive} (the one with the $5\sigma$ tension with Planck 2018 flat-$\Lambda$CDM \citep{aghanim2020planck})  shows no evidence of distance inconsistencies. This implies that under this test, we have no evidence of the tension coming from systematics.

\section*{Data availability}

The code required to reproduce the results in this article are available at \url{https://github.com/antonionajeraq/consistency_redshift_independent_galaxies.git}. The NED-D data is available at \url{https://ned.ipac.caltech.edu/Library/Distances/}, and the SH0ES 2022 data at \url{https://github.com/PantheonPlusSH0ES/DataRelease}.
% Other data underlying the article may be made available upon request to the authors.

\section*{Acknowledgements}

We thank Barry Madore, Claudia Maraston, Ian Steer, Richard Stiskalek and Tariq Yasin for useful inputs and discussion. JAN and HD are supported by a Royal Society University Research Fellowship (grant no. 211046).
% For the purposes of open access, the authors have applied a Creative Commons Attribution (CC BY) licence to any Author Accepted Manuscript version arising.

\bibliographystyle{mnras}
\bibliography{references}

\appendix

\begin{table*}
    \centering
        \caption{As Table~\ref{tab:results-standard-candles} but including only measurements that can be standardised through either $H_0$ or $\mu_\text{LMC}$.
        % Results for the standard candle distance indicators. We present the number of galaxies having at least two measurements by a given method, the number of measurement pairs in these galaxies, the KLD $p$-value and external consistency statistics. The $p$-value is shown in green if it is greater than 0.5 (indicating consistency), orange if it is between 0 and 0.05 (indicating inconsistency), and red if it is 0 (indicating severe inconsistency). The final column shows the fraction of other methods with which the method in question is externally consistent, out-of-the-box. %\hd{Can you put spaces around the slashes?}
        }
    \label{tab:results-standard-candles-calibrated}
    \begin{tabular}{cccccc}
    \hline
    Distance indicator & \# of galaxies & \# pairs & \# total measurements/references & KL $p$-value & External consistency \\
    \hline
CMD & 20 & 3454 & 1265 / 170 & \textcolor{OliveGreen}{0.9812} & 11 / 20 \\ 
\textbf{BL Lac Luminosity} & 2 & 2 & 20 / 11 & \textcolor{OliveGreen}{0.9215} & - \\ 
RR Lyrae & 8 & 38 & 26409 / 216 & \textcolor{OliveGreen}{0.8031} & 13 / 26 \\ 
TRGB & 2 & 2 & 1736 / 336 & \textcolor{OliveGreen}{0.4473} & 5 / 18 \\ 
RV Stars & 1 & 10 & 5 / 1 & \textcolor{OliveGreen}{0.087} & - \\ 
Cepheids & 36 & 1157 & 10956 / 325 & \textcolor{BrickRed}{0.0}  & 2 / 39 \\ 
GRB & 86 & 1214 & 729 / 17 & \textcolor{BrickRed}{0.0} & - \\ 
SNIa SDSS & 1849 & 7687 & 6054 / 1 & \textcolor{BrickRed}{0.0}  & 0 / 2 \\ 
Type II Cepheids & 2 & 13 & 62 / 22 & \textcolor{BrickRed}{0.0}  & 10 / 18 \\ 
SNIa & 4856 & 117978 & 30843 / 140 & \textcolor{BrickRed}{0.0}  & 7 / 25 \\ 
AGN time lag & 0 & 0 & 32 / 3 & - & 2 / 3 \\ 
AGB & 0 & 0 & 2 / 2 & - & 1 / 2 \\ 
B Stars & 0 & 0 & 2 / 2 & - & 2 / 3 \\ 
Black Hole & 0 & 0 & 17 / 2 & - & 2 / 2 \\ 
Blue Supergiant & 0 & 0 & 2 / 1 & - & 3 / 5 \\ 
BCG & 0 & 0 & 239 / 1 & - & 2 / 3 \\ 
Brightest Stars & 0 & 0 & 129 / 41 & - & 0 / 6 \\ 
Carbon Stars & 0 & 0 & 82 / 13 & - & 12 / 16 \\ 
Delta Scuti & 0 & 0 & 12 / 2 & - & 4 / 7 \\ 
FGLR & 0 & 0 & 7 / 7 & - & 0 / 3 \\ 
GCLF & 0 & 0 & 781 / 96 & - & 5 / 16 \\ 
GC SBF & 0 & 0 & 3 / 1 & - & 6 / 9 \\ 
HII LF & 0 & 0 & 17 / 2 & - & 0 / 4 \\ 
Horizontal Branch & 0 & 0 & 110 / 53 & - & 5 / 11 \\ 
\textbf{M Stars} & 0 & 0 & 6 / 5 & - & 2 / 6 \\ 
Miras & 0 & 0 & 36 / 27 & - & 8 / 15 \\ 
\textbf{Novae} & 0 & 0 & 18 / 11 & - & 11 / 17 \\ 
\textbf{OB Stars} & 0 & 0 & 4 / 3 & - & 6 / 7 \\ 
PNLF & 0 & 0 & 183 / 57 & - & 2 / 6 \\ 
Red Clump & 0 & 0 & 236 / 63 & - & 8 / 14 \\ 
RSV Stars & 0 & 0 & 9 / 6 & - & 3 / 8 \\ 
SX Phe Stars & 0 & 0 & 3 / 2 & - & 0 / 1 \\ 
Statistical & 0 & 0 & 292 / 27 & - & 14 / 37 \\ 
Subdwarf fitting & 0 & 0 & 1 / 1 & - & 0 / 1 \\ 
SZ effect & 0 & 0 & 283 / 19 & - & 4 / 5 \\ 
SBF & 0 & 0 & 1781 / 67 & - & 2 / 14 \\ 
SNII radio & 0 & 0 & 24 / 2 & - & 2 / 6 \\ 
White Dwarfs & 0 & 0 & 1 / 1 & - & 0 / 1 \\ 
Grav. Wave & 0 & 0 & 8 / 8 & - & 1 / 1 \\ 
\hline
    \end{tabular}
\end{table*}

\section{Results with Standardisable Measurements Only}
\label{sec:app}

Here we show the analogues of Tables \ref{tab:results-standard-candles} and \ref{tab:results-standard-rulers} but excluding measurements that do not list a $H_0$ or $\mu_\text{LMC}$ value and hence cannot be standardised.
% Thus, they can be calibrated to have $H_0 = 70$ km/s/Mpc and $\mu_\text{LMC} = 18.50$.
On the whole we find that the consistency little improves, indicating that the non-standardisability of many of the measurements was not a principal cause of inconsistency in the main text. For this reason we do not show the results after rescaling the uncertainties and/or shifting the means. We do however find that behaviour of the \texttt{CMD} indicator improves dramatically. This method also has a significant number of measurements, suggesting that it may be useful for constraining the distance ladder.
% This shows that a subset of measurements can be consistent, and thus provide reliable distances to different galaxies, making it suitable to cosmological analysis, like constraining $H_0$. 

\begin{table*}
    \centering
    \caption{As Table~\ref{tab:results-standard-rulers} but including only measurements that can be standardised through either $H_0$ or $\mu_\text{LMC}$.}
    \label{tab:results-standard-rulers-calibrated}
    \begin{tabular}{cccccc}
    \hline
    Distance indicator & \# of galaxies& \# pairs & \# total measurements/references & KL $p$-value & External consistency \\ \hline
SNII optical & 128 & 390 & 1622 / 89 & \textcolor{BrickRed}{0.0}  & 1 / 7 \\ 
Dwarf Galaxy Diameter & 0 & 0 & 16 / 1 & - & 6 / 9 \\ 
Eclipsing Binary & 0 & 0 & 273 / 48 & - & 9 / 15 \\ 
GC radius & 0 & 0 & 101 / 8 & - & 1 / 11 \\ 
G Lens & 0 & 0 & 86 / 14 & - & 0 / 1 \\ 
HII region diameter & 0 & 0 & 53 / 7 & - & 1 / 12 \\ 
\textbf{Maser} & 0 & 0 & 24 / 20 & - & 4 / 8 \\ 
Orbital Mech. & 0 & 0 & 3 / 3 & - & 0 / 2 \\ 
\textbf{Proper Motion} & 0 & 0 & 5 / 3 & - & 2 / 3 \\ 
Ring Diameter & 0 & 0 & 212 / 1 & - & 2 / 4 \\ 
\hline
GeV TeV ratio & 3 & 5 & 30 / 11 & \textcolor{RedOrange}{0.0124} & - \\ 
Faber-Jackson & 427 & 1907 & 1424 / 6 & \textcolor{RedOrange}{0.0016}  & 4 / 18 \\ 
D-Sigma & 548 & 2787 & 1995 / 9 & \textcolor{BrickRed}{0.0}  & 9 / 16 \\ 
FP & 807 & 1134 & 130214 / 22 & \textcolor{BrickRed}{0.0}  & 7 / 22 \\ 
IRAS & 404 & 639 & 2947 / 2 & \textcolor{BrickRed}{0.0}  & 3 / 28 \\ 
Tully-Fisher & 4972 & 99919 & 56436 / 78 & \textcolor{BrickRed}{0.0}  & 11 / 44 \\ 
Diameter & 0 & 0 & 6 / 2 & - & 9 / 15 \\ 
\textbf{GC FP} & 0 & 0 & 3 / 3 & - & 7 / 9 \\ 
H I + optical distribution & 0 & 0 & 1 / 1 & - & 0 / 1 \\ 
LSB galaxies & 0 & 0 & 2 / 1 & - & 1 / 4 \\ 
Magnitude & 0 & 0 & 111 / 3 & - & 1 / 9 \\ 
Sosies & 0 & 0 & 287 / 3 & - & 1 / 4 \\ 
Tertiary & 0 & 0 & 284 / 4 & - & 5 / 12 \\ 
Tully est & 0 & 0 & 1434 / 1 & - & 17 / 27 \\ 
\hline
    \end{tabular}

\end{table*}

\end{document}